\documentclass[12pt]{article}
\usepackage{amssymb}
\usepackage{amsfonts}
\usepackage{amsmath}
\usepackage{geometry}
\usepackage{accents}
\usepackage[onehalfspacing]{setspace}

\newtheorem{theorem}{Theorem}

\newtheorem{claim}{Claim}

\newtheorem{corollary}{Corollary}

\newtheorem{definition}{Definition}
\newtheorem{example}{Example}

\newtheorem{lemma}{Lemma}

\newtheorem{proposition}{Proposition}
\newtheorem{remark}{Remark}

\begin{document}

\title{A Dutch-book trap for misspecification\thanks{%
We would like to thank Pierpaolo Battigalli, Kyle Chauvin, Roberto Corrao,
Federico Echenique, Francesco Fabbri, Shaowei Ke, Jiangtao Li, Julien
Manili,\ Luciano Pomatto, Eran Shmaya, Marciano\ Siniscalchi, Micheal
Strevens, and Brad Weslake for the precious discussions on this project.
Emiliano Catonini gratefully acknowledges the financial support of the
National Science Foundation of China, Excellent Young Scientist Program
(overseas).}}
\author{Emiliano Catonini\thanks{%
New York University Shanghai, emiliano.catonini@nyu.edu.} and Giacomo Lanzani%
\thanks{%
UC Berkeley, giacomolanzani34@gmail.com.}}
\maketitle

\begin{abstract}
We provide Dutch-book arguments against misspecified Bayesian learning. An
agent progressively learns about a state and is offered a bet after every
discovery. We say the agent is deterministically Dutch-booked when they
would accept all bets, but their payoff is ex-post negative under each
state. More generally, we say that the agent is Dutch-booked when they would
accept all bets, but their expected payoff under each fundamental state is
negative. With this, the agent cannot be deterministically Dutch-booked if
and only if they update their beliefs using Bayes' rule, even with
misspecified likelihoods. The agent cannot be Dutch booked if and only if
they update their beliefs using Bayes' rule with a lexicographic prior and
using the correct data-generating process. We show that offers of financial
instruments and behavior in Monty Hall problems can be viewed as Dutch books
that extract a sure expected gain from a misspecified population.
\end{abstract}

\begin{quotation}
\end{quotation}

\newpage

\section{Introduction}

A growing literature (surveyed in Bohren and Hauser, 2025) documents the
consequences of updating beliefs using a misspecified model of the world.
Model misspecification leads entirely rational (i.e., Bayesian) agents
astray by inducing them to misinterpret the data and compute incorrect
posteriors. Is there a general sense in which model misspecification is
harmful for decision-making? We provide an answer to this question through a
two-fold Dutch-book argument.

The classical Dutch-book argument (Ramsey 1931, De Finetti 1937) is a
cornerstone of probability theory. It says that an agent (\emph{he}) may
accept a set of bets that returns an aggregate loss under every state (i.e.,
a Dutch book) if his preferences cannot be rationalized by one probabilistic
belief. Later (Lehman 1955, Kemeny 1955), the converse was also proven: an
agent cannot be Dutch-booked if he evaluates all bets with the same
probabilistic belief about the state. In the classical Dutch-book argument,
there is no time dimension. However, the agent may progressively receive
information about a fundamental state, and the bookmaker (\emph{she}) could
offer one bet conditional on every piece of news. Epistemologists have
studied this problem in a framework where the agent learns finer and finer
partitions of the state space on which he bets.\footnote{%
An exception is Pettigrew (2023): see the literature review.}\ Under the
assumption that every state is deemed possible at the outset, Teller (1973)
and Lewis (1999) have shown that an agent cannot be Dutch-booked if and only
if he updates beliefs by conditioning.\footnote{%
Rescorla (2022) surveys Dutch-book arguments and extends the argument for
conditioning to the presence of subjectively impossible states.}

However, in many problems of dynamic choice, the decision-relevant state
does not include the observed events; hence, learning is best represented as
the acquisition of noisy signals. That is the approach that is followed in
the entire economics literature on misspecified learning. In particular,
this literature employs the following learning environment, which we also
adopt here. There is a set of decision-relevant fundamental states and a set
of contingencies, partially ordered in time, in which the agent makes
choices. The fundamental states represent different (probabilistic) DGPs for
contingencies. Thus, the contingencies have objective probabilities
conditional on the fundamental states. Then, the question arises: what if
the agent is (Bayesian but) misspecified about these probabilities?

We call a \emph{deterministic Dutch} \emph{book} a system of bets on the
fundamental state, one for each contingency, so that the \emph{cumulative}
payoff of the agent is nonpositive under all fundamental states and paths,
and negative under some fundamental state and path.\footnote{%
Thus, the notion of deterministic Dutch book extends to this paper's
framework the notion of \textquotedblleft diachronic Dutch
book\textquotedblright\ employed by the arguments for conditioning.} More
generally, we call a system of bets a \emph{Dutch book} if it yields a
nonpositive \emph{expected} payoff under every fundamental state and a
negative expected payoff under some fundamental state, where the
expectations are computed with the probabilities of the contingencies as
determined by the fundamental state under consideration. An agent is
(deterministically) Dutch-booked when she is willing to accept all the bets
in a (deterministic) Dutch book.\footnote{%
We use the following tie-breaking rule:\ the agent does not accept bets with
zero subjective expected payoff.} In other words, a deterministic Dutch book
guarantees that, under every possible DGP, the agent surely receives a
nonpositive payoff and possibly a strictly negative payoff. Instead, a Dutch
book requires that the agent's expected payoff, given the probabilities over
contingencies induced by every DGP, is nonpositive and strictly negative for
some DGP.

First, we show that an agent cannot be deterministically Dutch-booked if and
only if his beliefs are \emph{forward-consistent}, i.e., the agent updates
his beliefs using Bayes rule whenever possible. Since forward-consistency
does not require the use of the correct conditional probabilities of the
contingencies, agents that are Bayesian but misspecified about the \emph{%
relative probabilities in the likelihood function} (i.e., they may apply
Bayes rule to the wrong likelihood function linking states and the
contingencies) are immune to deterministic Dutch books. However, an agent
with the more radical form of misspecification, in which \emph{a state is
subjectively believed possible given a contingency objectively impossible
under that state}, can be deterministically Dutch-booked. Moreover, forward
consistency is equivalent to Bayes-plausibility (Kamenica and Gentzkow,
2011), i.e., the requirement that each belief is a convex combination of the
beliefs that immediately follow. Thus, Bayes-plausibility is the coherence
rule among beliefs in counterfactual contingencies that characterizes an
agent who cannot be deterministically Dutch-booked.\footnote{%
Note that the necessity of forward-consistency to avoid Dutch-booking does
not follow from the existing results that establish the necessity of
conditioning in the partitional framework, because of the multiplicity of
learning paths that are consistent with the same state. Thus, given a
violation of forward-consistency between a contingency and its successors,
the bookmaker needs to find an acceptable bet for the first contingency such
that the opposite bet will be accepted in each of the following
contingencies (despite the potentially different posteriors), and not just
in one as in the partitional framework. Another way of seeing this
difficulty is that, in this paper, the bookmaker is constrained to offer
bets on only one dimension of the \textquotedblleft grand state
space,\textquotedblright\ which also includes the learning paths. The
solution to this problem will be an application of Farkas lemma.}

Second, and most importantly, we show that a Bayesian agent cannot be
Dutch-booked (\textquotedblleft in objective expected
terms\textquotedblright ) if and only if he updates beliefs from a
lexicographic prior with the correct likelihood function. This is perhaps
surprising: since there is no \textquotedblleft bad prior\textquotedblright\
over fundamental states\ (because there is no objective one), it is hard to
conceive in what sense updating the prior with wrong likelihoods could
generate a \textquotedblleft bad posterior\textquotedblright\ over
fundamental states. Our Dutch-book argument offers a precise sense in which
model misspecification is harmful for decision-making: a misspecified agent
could accept sure expected losses and forfeit sure expected wins. The
argument also offers a motivation for having a prior, even when there is no
decision to take before the arrival of information. The use of a \emph{%
lexicographic} prior (in place of free belief revision after each surprise)
is crucial to guarantee the coherence of beliefs across counterfactual
contingencies, thereby shielding the agent from a Dutch book. We call a
belief system \emph{completely consistent }(henceforth CCBS) when it is
derived from a subjective lexicographic prior with the correct likelihood
function. More generally, the argument justifies having coherent beliefs
that can be rationalized with a prior, although the agent may not truly have
one; rather, the agent may reason coherently after different discoveries. We
call \emph{complete coherence} the rule of coherence between beliefs in
counterfactual contingencies that shelters the agent from Dutch-booking. The
rule is easy to interpret; it is based on the idea that differences in the
subjective odds ratios of states across contingencies shall only reflect
differences in the objective odds ratios of the contingencies given the
states.

Importantly, we also show that the possibility of Dutch-booking a
misspecified agent does not rely on the bookmaker knowing the true data
generating process perfectly. As long as a well defined measure of
discrepancy between the agent and bookmaker beliefs is sufficiently large
compared to the bookmaker misspecification, a Dutch book can be constructed.

Leveraging the separation between subjective (over fundamental states) and
objective (given the fundamental state) uncertainty, our general notion of
Dutch book aggregates payoffs across counterfactual contingencies according
to the objective probability over them prescribed by the fundamental state.
Besides being a useful theoretical tool to uncover a form of
\textquotedblleft irrationality\textquotedblright , this weaker form of
Dutch book could also be of applied relevance, even though a single agent
will only find himself in one of many counterfactual contingencies. First, a
large population of agents making choices in different contingencies will
realize the aggregate payoff under the actual state. So, our Dutch-book
argument suggests a disadvantage for \textquotedblleft
divided\textquotedblright\ populations, in which different agents hold
different (prior) beliefs and can thus be exploited in different situations.
Such a divided population naturally emerges in a repeated setting when
agents do not share their experiences, suggesting an important role for
aggregating feedback. Second, a single agent may perish in the long run
after being exposed to i.i.d. occurrences of the same dynamic decision
problem. Third, an agent may be offered the whole set of conditional bets at
the ex-ante stage, as in a speculative-trade context (see the related
literature below). Traditional Dutch-book arguments are often criticized for
the possibility that the agent would realize being Dutch-booked when all
bets are offered simultaneously. However, in our case, a Dutch-booked agent
will still compute a positive expected payoff under some state by using a
misspecified likelihood function.\footnote{%
Dutch-book arguments have also been criticized for the need of a bookmaker
who knows the agent's beliefs. We disagree with this critique, as a
bookmaker may find a Dutch book that the agent wants to accept by trial and
error. In a financial market, the agent may also self-pick a set of assets
that jointly constitute a Dutch book, without the presence of a strategic
bookmaker.}

We then apply our theorems to one commonly studied form of misspecification,
correlation neglect. Since agents with this bias update beliefs using Bayes
rule, our first result implies that they are shielded from deterministic
Dutch books. However, they use an incorrect likelihood that neglects the
correlation between contingencies, thereby exposing them to general Dutch
books, consistent with our second result. In particular, we illustrate the
nature of those Dutch books in a venture capital setting and in a variation
of the Monty Hall game, with secretly assigned doors. In the former
application, the Dutch book consists of the venture capitalist offering
different combinations of convertible debt and ownership purchases depending
on whether initial milestones are achieved. In the second case, the Dutch
book offers terms that seem inviting if the remaining doors were equally
likely, but that induce sure expected loss under the mechanism used to
determine what door is open initially.

\paragraph{Related literature}

In the context of sequential games, Siniscalchi (2022)\ introduces the
notion of \emph{structural rationality}, which is based on the use of CCBSs
over the opponents' strategies.\footnote{%
Interestingly, structural rationality is motivated by the inability of
sequential rationality to capture coherent conditional preferences that can
be derived by updating a prior preference relation. CCBSs guarantee that all
conditional preferences can be derived from a prior lexicographic preference
relation. In other words, they guarantee a stronger form of Bayesianism,
which might be called \textquotedblleft Bayesian updating of
preferences\textquotedblright\ rather than \textquotedblleft Bayesian
updating of beliefs.\textquotedblright} In Siniscalchi (2022), as well as
Battigalli et al. (2023), a CCBS is defined as an array of beliefs, one for
each information set, that is consistent with a \emph{Complete} Conditional
Probability System (CCPS): an array of probability measures, one for each
subset of opponents' strategies, disciplined by the chain rule of
probability.\footnote{%
Conditional probability systems were first introduced by Renyi (1955). CCPSs
were introduced in the analysis of games by\ Myerson (1986).} In turn, a
CCPS can be derived from a lexicographic probability system by conditioning.
In a sequential game, a player progressively rules out opponents' strategies
by observing their moves; that is, learning occurs in a partitional fashion.
So, Bayes rule boils down to conditioning, and hence the definition of CCBS
in Siniscalchi (2022) and Battigalli et al. (2023) is a special case of the
definition of this paper. Thus, the Dutch-book argument for complete
consistency implies that the standard notion of a conditional probability
system defined on the collection of \emph{observable} events falls short of
the level of coherence that shelters from Dutch-booking in objective
expected terms. The working paper of Siniscalchi (2022) also characterizes
CCBSs by generalizing the chain rule to apply to non-nested conditional
events. In the Appendix, we provide a direct proof of the equivalence
between the generalized chain rule and our coherence rule in sequential
games.

Pettigrew (2023) considers an agent with a credence at time $t$ and a vector
of possible credences at $t^{\prime }>t$, and shows that the agent can be
Dutch-booked if and only if his credence at $t$ is not in the convex hull of
the credences at $t^{\prime }$ (which he calls \textquotedblleft General
Reflection Principle\textquotedblright ). This result yields our equivalence
between Bayes plausibility and the impossibility of deterministic
Dutch-booking for the special case of a tree with depth $1$ and discoveries
that do not rule out any state.\footnote{%
When some discoveries objectively rule out some states, statement II.a of
Theorem 1 in Pettigrew (2023) fails. From his theorem, Pettigrew also
derives an argument for conditioning in the partitional framework. In our
richer framework, we can derive our more general argument for forward
consistency, establishing the impossibility of achieving a deterministic
Dutch-booking against likelihood misspecification.}

Chen et al. (2015) show that the posteriors of different agents, who update
a common prior based on different discoveries, are common knowledge if and
only if they cannot be Dutch-booked in aggregate.\footnote{%
See also Lehrer and Samet (2011) for an alternate characterization of
\textquotedblleft agreeing to agree\textquotedblright\ between only two
agents.} Their notion of a Dutch book considers only \emph{realized}
discoveries and sums up the payouts under each state. This amounts to
assuming that the bookmaker can choose the bets based on the entire set of
realized discoveries. So, when the agents have different (and thus not
commonly known) posteriors, the bookmaker can exploit an informational
advantage over each agent, who is unaware of the other agents' discoveries.
In other words, the bookmaker can make a profit from arbitrage. This is
ruled out by our notion of a Dutch book, in which each possible discovery is
associated with a bet, independently of what other discoveries (using the
population interpretation) may have been made by other agents. This
reconciles the possibility of Dutch-booking Bayesian agents in Chen et al.
(2015) with the impossibility results of this paper.

Molavi (2025) shows that beliefs are consistent with Bayesian updating if
and only if the mean posterior is absolutely continuous with respect to the
prior. This finding generalizes the earlier work of Shmaya and Yariv (2016)
to the case in which the decision maker's subjective beliefs may not have
the same support as the true data-generating process. Fudenberg and Lanzani
(2026) characterize when beliefs are consistent with Bayesian updating from
a conditional i.i.d. model.

There are several related contributions in the literature on misspecified
Bayesian learning, which is excellently surveyed in Bohren and Hauser (2025)
and Esponda and Pouzo (2026). Here we discuss the part of this literature
concerned with measuring the loss induced by misspecification and how to
avoid these losses. Lanzani (2025) characterizes a behavior rule that
prevents misspecification from damaging the agent beyond his maxmin payoff,
preserving consistency when correctly specified. Frick et al. (2024) provide
a rank of forms of misspecification on the basis of the damages they induce,
and Gossner and Steiner (2018) provides a measure of cost of misperceiving
the payoff function. Bohren and Hauser (2024) characterize the conditions
under which a departure from Bayesian updating (e.g., underinference from
signals) can be rationalized as a consequence of Bayesian updating within a
misspecified model. Cerreia-Vioglio et al. (2025) show that misspecification
in coordination games can be highly damaging, even for agents concerned
about it. Compared to this literature, we provide a novel reason for why
some misspecification should or should not be expected to persist based on
their susceptibility to Dutch books. Moreover, as our application to
correlation neglect clearly shows, the form of misspecification that are
susceptible to Dutch book significantly differ from the one that have been
deemed unstable due to light-bulb realizations or evolutionary pressure
(see, Gagnon-Bartsch et al. 2023, Fudenberg and Lanzani 2023, Ba 2026, He
and Libgober 2025).

Finally, our results are similar in spirit to those in the no-trade
literature, the closest being in Morris (1994). Morris considers agents who
trade claims over commodities that are contingent on a payoff-relevant state
and on the realization of a signal. In the case of two traders, a single
money commodity, and risk neutrality, Morris' model coincides with ours with
a tree of depth $1$. Assuming that the traders have no ex-ante incentive to
exchange claims that are contingent on the state only (which in our setting
boils down to having the same prior), Theorem 2.1 in Morris (1994) shows
that there exists trade that gives a positive expected payoff to each trader
if and only if they disagree on the likelihood function mapping states into
probabilities over signals. Our main theorem further shows (in a more
general learning framework) that such trade can take a very specific form,
whereby one trader has a positive expected payoff conditional on each state,
and the other has a positive expected payoff conditional on each signal.
Suppose the first trader (the bookmaker) is using the correct likelihood
function. In that case, the second trader (our agent) gets an \emph{%
objectively} negative expected payoff conditional on each state, and thus is
Dutch-booked. Conversely, we show that the existence of such trade
guarantees that the disagreement between the two traders \emph{must} be
about the likelihood function, not (just) the prior.

\section{Framework}

There is a finite set $\mathcal{S}$ of possible (fundamental) \emph{states}.
We describe the possible learning processes through an arborescence (i.e., a
set of trees) of \emph{contingencies} $\mathcal{H}$. With this, we are not
assuming the existence of just one initial contingency in which the agent
forms beliefs before learning anything about the state. A \emph{learning path%
} is a path of contingencies from an \textquotedblleft
initial\textquotedblright\ contingency that is not preceded by any other
contingency (i.e., the root of a tree), to a \textquotedblleft
terminal\textquotedblright\ contingency that is not followed by any other
contingency (i.e., a leaf of that tree). Let $\mathcal{L}$ denote the set of
possible learning paths and $L(h)$ the set of paths that go through a
contingency $h$, and $H(l)$ the set of contingencies through which learning
path $l$ goes. Each state $s\in \mathcal{S}$ induces an objective
distribution $\eta ^{s}\in \Delta (\mathcal{L})$ over learning paths, and
thus states can be interpreted as different possible data-generating
processes. Let $L(s)=\mathrm{supp}\eta ^{s}$. For each $h\in \mathcal{H}$,
let $p(h|s)=\sum_{l\in L(h)}\eta ^{s}(l)$ denote the probability of reaching
contingency $h$ under state $s$. Let $S(h)$ denote the set of states $s$
that are consistent with contingency $h$, that is, $p(h|s)>0$. Without loss
of generality, we assume that every contingency is consistent with some
state. An important special case is the one in which the probability of a
contingency $h$ is the same across all the consistent states $s$, that is, $%
p(h|s)=p(h|s^{\prime })=:p^{h}$ for all $h\in \mathcal{H}$ and $s,s^{\prime
}\in S(h)$. We will refer to this as the \textquotedblleft independence
case\textquotedblright .

We now introduce an example that we will repeatedly use to illustrate our
contribution.

\begin{example}
Larry wants to go out and watch the New Year's fireworks, but the
municipality does not disclose their location anymore to avoid crowding.
There are three possible locations: the main square, the marina, and the
central park, $\mathcal{S=}\left\{ sq,ma,pa\right\} $. Three bus lines run
on his street: the red line goes to the square, the blue to the marina, and
the green to the park. Then, after taking a bus (say the red bus), Larry may
find himself in one of two contingencies: the one in which he arrives at the
correct location ($sq$) and the one in which the fireworks are elsewhere ($%
mp $). So, without introducing an initial contingency, we have $\mathcal{H=}%
\left\{ sq,ma,pa,sm,mp,ps\right\} $, where each contingency $h$ is labeled
with the only state or with the initials of the two states $S(h)$ that are
consistent with it. Since the contingencies are all unordered, the learning
paths have length $1$ and coincide with them.

Larry is late, so his only chance to see the fireworks is to take the first
bus that arrives, regardless of the line, and hope it goes to the correct
location. Suppose that the three lines arrive in the given order at constant
time intervals, independently of the state. Then, each state $s\in \mathcal{S%
}$ induces a uniform distribution $\eta ^{s}$ over the three learning
paths/contingencies $h$ such that $s\in S(h)$, and we are in the
independence case: for each $h\in \left\{ sm,mp,ps\right\} $ and $%
s,s^{\prime }\in S(h)$, $p(h|s)=p(h|s^{\prime })=1/3$.\hfill $\triangle $
\end{example}

\section{Belief systems}

We assume that, in each contingency $h$, the agent has a probabilistic
belief $\mu (\cdot |h)\in \Delta (\mathcal{S})$ such that $\mu (S(h)|h)=1$%
--- we call $(\mu (\cdot |h))_{h\in \mathcal{H}}$ a \textbf{belief system}.
This maintained assumption, besides requiring probabilistic beliefs at any
point in time, rules out the most severe forms of model misspecification in
which an agent assigns positive probability to states that have become
objectively impossible.

We say that a belief system is \emph{forward-consistent} if the agent's
beliefs are consistent with Bayesian updating from his previous belief using
some (conditional) likelihood function, whenever possible. Observe that such
a conditional likelihood function is not required to be the objective one.

\begin{definition}
\label{Def: FCBS}A belief system $(\mu (\cdot |h))_{h\in \mathcal{H}}$ is 
\textbf{forward-consistent} if, for every non-terminal $h\in \mathcal{H}$,
calling $h^{1},...,h^{n}$\ its immediate successors, there exist conditional
probability distributions 
\begin{equation*}
\left( \widetilde{p}(\cdot |h,s)\right) _{s\in S(h)}\in \left( \Delta
(\left\{ h^{1},...,h^{n}\right\} )\right) ^{S(h)}
\end{equation*}%
such that, for each $m\in \left\{ 1,...,n\right\} $, if $\widetilde{p}%
(h^{m}|h,s)>0$ for some $s\in \mathrm{supp}\mu (\cdot |h)$, then $\mu (\cdot
|h^{m})$ can be derived with Bayes rule from $\mu (\cdot |h)$ using $(%
\widetilde{p}(h^{m}|h,s))_{s\in S(h)}$.
\end{definition}

It is well known that beliefs are consistent with this form of updating if
and only if the \textquotedblleft prior belief\textquotedblright\ (i.e., the
belief at $h$) is a convex combination of the \textquotedblleft posterior
beliefs\textquotedblright\ (i.e., the beliefs at $h^{1},...,h^{n}$). (The
Appendix contains a formal proof of this fact.)

\begin{definition}
A belief system $(\mu (\cdot |h))_{h\in \mathcal{H}}$ is \textbf{%
Bayes-plausible} if, for every non-terminal $h\in \mathcal{H}$ with
immediate successors $h^{1},...,h^{n}$, there exist $\alpha _{1},...,\alpha
_{n}\in \left[ 0,1\right] $ such that 
\begin{equation*}
\sum_{m=1}^{n}\alpha _{m}=1\quad \text{and\quad }\sum_{m=1}^{n}\alpha
_{m}\mu (\cdot |h^{m})=\mu (\cdot |h).
\end{equation*}
\end{definition}

As we will show, Bayes-plausibility is the minimal requirement of coherence
among beliefs that shields against deterministic Dutch-booking.

\bigskip

We will define a Completely Consistent Belief System as a belief system that
can be derived from a Lexicographic Conditional Probability System.

\begin{definition}[Blume et al., 1991]
\label{Def: LCPS}A list of probability measures $\bar{\mu}=(\mu ^{1},...,\mu
^{n})$ over $\mathcal{S}$ is a \textbf{Lexicographic Conditional Probability
System} if for every $s\in \mathcal{S}$, there exists exactly one $m\in
\left\{ 1,...,n\right\} $ such that $\mu ^{m}(s)>0$.
\end{definition}

Definition \ref{Def: LCPS} embodies a \textquotedblleft
full-support\textquotedblright\ requirement:\ every state is given positive
probability at some level of the LCPS. In this way, from every LCPS, one can
always derive a belief system as follows.

\begin{definition}
\label{Def: CCBS}A belief system $(\mu (\cdot |h))_{h\in \mathcal{H}}$ is 
\textbf{completely consistent} if there exists a LCPS $\bar{\mu}=(\mu
^{1},...,\mu ^{n})$ such that, for each $h\in \mathcal{H}$, $\mu (\cdot |h)$
can be derived with Bayes rule using $\left( \eta ^{s}\right) _{s\in 
\mathcal{S}}$ from the earliest $\mu ^{m}$ such that $\mu ^{m}(S(h))>0$,
i.e., $m=\min \left\{ k:\mu ^{k}(S(h))>0\right\} $ and%
\begin{equation*}
\mu (s|h)=\frac{p(h|s)\mu ^{m}(s)}{\sum_{s^{\prime }\in S}p(h|s^{\prime
})\mu ^{m}(s^{\prime })}\qquad \forall s\in S(h)\text{.}
\end{equation*}
\end{definition}

In a CCBS, given each contingency $h$, the belief $\mu (\cdot |h)$ is
derived with Bayes rule from the first theory in the LCPS that can explain $%
h $, i.e., that assigns positive probability to some $s\in S(h)$.\footnote{%
Ortoleva (2012) proposes and axiomatizes a generalization of LCPS that
allows the agent to change prior after small but positive probability event
in a way that better matches the empirical evidence about belief revisions.}
Moreover, Bayes rule must be applied using the objective likelihood $p$.
Note that, for any two contingencies $h,h^{\prime }$ such that$\
S(h)=S(h^{\prime })$, the beliefs derived from the LCPS with Bayes rule may
differ. However, they are derived from the same measure in the LCPS. This is
because a contingency may convey more information than the set of consistent
states, as it may be reached with different probabilities under different
states. When this is not the case, i.e., in the independence case, Bayes
rule boils down to conditioning and Definition \ref{Def: CCBS} coincides
with the definition of CCBS in Siniscalchi (2022) and Battigalli et al.
(2023) --- see the Appendix for details.

\bigskip

We will characterize CCBSs in terms of coherence among the subjective
relative probabilities of states in different contingencies. However,
because the objective probabilities of contingencies may differ across
states, these comparisons are meaningful only after discounting the
probability attributed to a state in a contingency by the probability that
the state induces the contingency.

\begin{definition}
\label{Def: odds ratio}Given a belief system $(\mu (\cdot |h))_{h\in 
\mathcal{H}}$, a contingency $h$, and a pair of states $s,s^{\prime }\in
S(h) $ such that $\mu (s|h)+\mu (s^{\prime }|h)>0$, the (possibly infinite)\ 
\textbf{discounted} \textbf{odds ratio} between $s$ and $s^{\prime }$ at $h$
is%
\begin{equation*}
o(s,s^{\prime }|h)=\frac{\mu (s|h)}{p(h|s)}\frac{p(h|s^{\prime })}{\mu
(s^{\prime }|h)}.
\end{equation*}

A \textbf{generalized odds ratio} between $s$ and $s^{\prime }$ is the
product of a finite concatenation of discounted odds ratios%
\begin{equation*}
o(s,s^{1}|h^{1})\cdot o(s^{1},s^{2}|h^{2})\cdot ...\cdot o(s^{n-1},s^{\prime
}|h^{n})
\end{equation*}%
such that, if one discounted odds ratio is $0$, no discounted odds ratio is
infinite.
\end{definition}

A discounted odds ratio compares the probabilities attributed to two states
in the same contingency. A generalized odds ratio compares the probabilities
of two states, $s$ and $s^{\prime }$, in two different contingencies, $h^{1}$
and $h^{n}$, via a chain of comparisons with other states ($s$ with $s^{1}$, 
$s^{1}$ with $s^{2}$, ..., until $s^{n-1}$ with $s^{\prime }$) in different
contingencies ($h^{1}$ for $s,s^{1}$, $h^{2}$ for $s^{1},s^{2}$, and so on).%
\footnote{%
The formal definition allows contingencies and states to repeat, because
this will be convenient for the proofs.} In the independence case,
discounted odds ratios boil down to (standard)\ odds ratios.

The strong rule of coherence that shelters from Dutch-booking in objective
expected terms is the equivalence of all generalized odds ratios between any
two states.

\begin{definition}
We say that a belief system is \textbf{completely coherent} when, for every $%
s,s^{\prime }\in \mathcal{S}$, the generalized odds ratios between $s$ and $%
s^{\prime }$ are all identical.
\end{definition}

We illustrate the notions of belief system and generalized odds ratio
through the running example.

\begin{example}
\label{Ex:Larry2}To obtain a CCBS, let us fix a LCPS $\bar{\mu}=(\mu
^{1},...,\mu ^{n})$. Because each contingency has the same probability under
the two consistent states (i.e., we are in the independence case), Bayes
rule reduces to the chain rule, and discounted odds ratios reduce to odds
ratios. So, from $\bar{\mu}$, we can derive a belief system $\mu =\left( \mu
(\cdot |h)\right) _{h\in \mathcal{H}}$ as follows. For each $h\in \left\{
sq,ma,pa\right\} $, $\mu (s|h)=1$ for $s=h$. Next, suppose first that $\mu
^{1}$ assigns positive probability to at least $2$ states. Then, for each $%
h\in \left\{ sm,mp,ps\right\} $, $\mu ^{1}(S(h))>0$, and we get%
\begin{equation*}
\mu (s|h)=\frac{\mu ^{1}(s)}{S(h)}\qquad \forall s\in S\left( h\right) .
\end{equation*}%
If instead $\mu ^{1}$ assigns probability $1$ to a state $\bar{s}$, we
obtain $\mu (\bar{s}|h)=1$ whenever $\bar{s}\in S(h)$, and the belief in any
contingency $h$ such that $\bar{s}\notin S(h)$ is given by%
\begin{equation*}
\mu (s|h)=\frac{\mu ^{2}(s)}{\mu ^{2}(S(h))}\qquad \forall s\in S\left(
h\right) .
\end{equation*}%
Given any two states $s,s^{\prime }$, one can compute an odds ratio between $%
s$ and $s^{\prime }$ in the only contingency $h$ such that $S(h)=\left\{
s,s^{\prime }\right\} $; call it $o(s,s^{\prime })$. Then, one can also
compute a generalized odds ratio between $s$ and $s^{\prime }$ as $%
o(s,s^{\prime \prime })\cdot o(s^{\prime \prime },s^{\prime })$ ($s^{\prime
\prime }\notin \left\{ s,s^{\prime }\right\} $), unless one is zero and the
other infinite. For instance, when $\mu ^{1}(ma)\in (0,1)$, we can compute 
\begin{eqnarray*}
o(sq,ma)\cdot o(ma,pa) &=&\frac{\mu (sq|sm)}{\mu (ma|sm)}\cdot \frac{\mu
(ma|mp)}{\mu (pa|mp)}=\frac{\mu ^{1}(sq)}{\mu ^{1}(ma)}\cdot \frac{\mu
^{1}(ma)}{\mu ^{1}(pa)} \\
&=&\frac{\mu ^{1}(sq)}{\mu ^{1}(pa)}=\frac{\mu (sq|ps)}{\mu (pa|ps)}%
=o(sq,pa),
\end{eqnarray*}%
which can be infinite if $\mu ^{1}(pa)=0$ and thus $\mu ^{1}(sq)>0$, or
zero, vice versa. Note the equality between the odds ratio and the
generalized odds ratio.

Suppose now that, after not finding the fireworks, Larry regrets not going
down to the bus stop a little earlier and believes that, most likely, the
previous bus would have brought him to the right location. In this case, he
could have the following belief system $\hat{\mu}$:%
\begin{eqnarray*}
\hat{\mu}(sq|sq) &=&\hat{\mu}(ma|ma)=\hat{\mu}(pa|pa)=1, \\
\hat{\mu}(sq|ps) &=&\hat{\mu}(ma|sm)=\hat{\mu}(pa|mp)=3/4\text{.}
\end{eqnarray*}%
Under this belief system, the odds ratio and the generalized odds ratio
between any two states differ. For instance, in contingency $ps$ we have $%
o(sq,pa)=3$, but in contingency $sm$ we have $o(sq,ma)=1/3$ and in
contingency $mp$ we have $o(ma,pa)=1/3$, therefore we can compute a
generalized odds ratio between $sq$ and $pa$ of $1/9$ instead of $3$. Thus, $%
\hat{\mu}$ is not completely coherent.

Every belief system is trivially forward-consistent, as every path consists
of just one (terminal)\ contingency. If we introduce a belief of Larry
before taking the bus, i.e., an initial contingency $h^{0}$, then
Bayes-plausibility is guaranteed by the condition that for each $h\in
\left\{ sq,ma,pa\right\} $, $\mu (h|h)=1$, as every initial belief would
trivially be a convex combination of these three vertices of $\Delta \left( 
\mathcal{S}\right) $. Thus, every belief system would still be
forward-consistent.\hfill $\triangle $
\end{example}

\section{Dutch books and Dutch-booking}

Traditionally, a Dutch book is a set of bets such that, no matter the
realization of the state, the cumulative payoff from the bets is never
positive and sometimes negative for the gambler (thus never negative and
sometimes positive for the bookmaker). An agent is Dutch-booked if he
accepts all the bets in a Dutch book. Our agent, however, would never accept
a classical Dutch book if all the bets are proposed simultaneously, as he
has a well-defined probabilistic belief at any point in time. We now
investigate whether, instead, a system of gambles proposed in different
contingencies can induce the same kind of state-wise losses.\footnote{%
Our agent's risk neutrality and probabilistic sophistication at every
contingency imply that, when he is willing to accept a bet, he is also
willing to accept the sum of the bets. Thus, we can assume, without loss of
generality, that our agent is offered only one bet in each contingency.}

\begin{definition}
\label{Def: system of gambles}A \textbf{gamble} is a map $\gamma :\mathcal{S}%
\rightarrow 
\mathbb{R}
$ that specifies, for each $s\in \mathcal{S}$, the gain or loss $\gamma (s)$
for the agent if $s$ realizes. A \textbf{system of gambles} is an array $%
g=(g(\cdot |h))_{h\in \mathcal{H}}$ where, for each $h\in \mathcal{H}$, $%
g(\cdot |h)$ is a gamble.
\end{definition}

What does it mean, for a system of gambles, to be a \textquotedblleft Dutch
book\textquotedblright ? An intuitive notion is that the bookmaker obtains a
never negative and sometimes positive profit under every state and path. We
call this notion \textquotedblleft deterministic Dutch
book\textquotedblright . But the existence of objective conditional
probabilities of paths allows us to introduce also a more permissive notion
of Dutch book, whereby the bookmaker obtains a non-negative/positive profit
conditional on each fundamental state only in objective expected terms.

\begin{definition}
\label{Def: Dutch book}A system of gambles $g=(g(\cdot |h))_{h\in \mathcal{H}%
}$ is a \textbf{Dutch book} if for every $s\in \mathcal{S}$,%
\begin{equation}
\sum_{h\in \mathcal{H}}p(h|s)g(s|h)\leq 0,  \label{Eq: DB1}
\end{equation}%
and for some $s\in \mathcal{S}$,%
\begin{equation}
\sum_{h\in \mathcal{H}}p(h|s)g(s|h)<0.  \label{Eq: DB2}
\end{equation}%
A system of gambles $g=(g(\cdot |h))_{h\in \mathcal{H}}$ is a \textbf{%
deterministic Dutch book }if%
\begin{equation}
\sum_{h\in H(l)}g(s|h)\leq 0,\qquad \forall s\in S,\forall l\in L\left(
s\right)  \label{Eq: DDB}
\end{equation}%
and for some $s\in \mathcal{S}$ and $l\in L\left( s\right) $,%
\begin{equation}
\sum_{h\in H(l)}g(s|h)<0.  \label{Eq: strict deterministic}
\end{equation}
\end{definition}

Note that a deterministic Dutch book does not depend on the objective
probabilities of paths, hence it is a Dutch book no matter what such
probabilities are. Moreover, Dutch books, deterministic or not, do not
depend on the objective probabilities of the states.

Since each gamble is proposed to the agent in a specific contingency, the
acceptance of the gamble will depend on the contingent belief.

\begin{definition}
\label{Def: willingness to accept}Fix a probability measure $\nu \in \Delta (%
\mathcal{S})$ and a gamble $\gamma $. We say that the agent is \textbf{%
willing to accept }$\gamma $ given $\nu $ if%
\begin{equation*}
\sum_{s\in \mathcal{S}}\nu (s)\gamma (s)>0\text{,}
\end{equation*}%
or $\gamma (s)=0$ for every $s\in \mathcal{S}$.

We say that an agent with belief system $(\mu (\cdot |h))_{h\in \mathcal{H}}$
is willing to accept a system of gambles $g=(g(\cdot |h))_{h\in \mathcal{H}}$
if, for every $h\in \mathcal{H}$, he is willing to accept $g(\cdot |h)$
given $\mu (\cdot |h)$.

We say that the agent is \textbf{(deterministically) Dutch-booked} if he is
willing to accept a (deterministic)\ Dutch book.
\end{definition}

Definition \ref{Def: willingness to accept} captures the following
tie-breaking rule: our agent declines gambles that yield zero subjective
expected payoff (except for the trivial \textquotedblleft
null\textquotedblright\ gamble, as the bookmaker is free to offer no
gamble). This rules out in a simple way the uninteresting case in which the
agent can be Dutch-booked by just charging negative payoffs for the states
he subjectively deems impossible.

To conclude this section, we illustrate Dutch-booking in the running example.

\begin{example}
\label{Ex:Larry3}A bookmaker follows Larry and, in case the bus does not
arrive at the correct location, she is ready to propose a bet on the
fireworks' location. If the bus goes to the marina, she is thinking of
offering a bet that yields $9$ if the fireworks are at the square and $-10$
if they are at the park; likewise at the square and at the park. Altogether,
we have the following system of bets:%
\begin{eqnarray*}
\hat{g}(sq|ps) &=&\hat{g}(ma|sm)=\hat{g}(pa|mp)=9; \\
\hat{g}(pa|ps) &=&\hat{g}(sq|sm)=\hat{g}(ma|mp)=-10; \\
\hat{g}(s|h) &=&0\qquad \forall h\in \left\{ sq,ma,pa\right\} .
\end{eqnarray*}%
Under every state $s$, we have%
\begin{equation*}
\sum_{h\in \mathcal{H}}p(h|s)\hat{g}(s|h)=\frac{1}{3}(9-10+0)=-\frac{1}{3}<0,
\end{equation*}%
so, this is a Dutch book.

If the bookmaker can also propose a bet to Larry before the bus arrives, she
may try to deterministically Dutch-book him. However, for him to accept this
bet in contingency $h^{0}$, she would need to offer a positive prize under
at least one state $s$.\ But then, to obtain a deterministic Dutch book, she
would need to offer a negative prize for $s$ in contingency $h=s$. Under the
imposed condition that Larry \textquotedblleft believes what he
knows\textquotedblright\ (i.e., $\mu (s|h)=1$), Larry would reject the bet.
Thus, Larry cannot be deterministically Dutch-booked. As already observed in
Example \ref{Ex:Larry2}, the same condition also guarantees Bayes
plausibility/forward consistency. This is not a coincidence, and the
argument is general: If, in a tree of depth 2, the agent has the possibility
of learning the state (whichever it is), his beliefs are always
forward-consistent, and he cannot be deterministically Dutch-booked. In the
next section, we will show that the tight connection between deterministic
Dutch-booking and forward-consistency extends beyond this special
case.\hfill $\triangle $
\end{example}

\section{The Dutch-book theorem for forward-consistency\label{Section:
forward}}

We now establish the equivalence among forward consistency, Bayes
plausibility, and immunity from deterministic Dutch books. The argument that
forward-consistency shelters against deterministic Dutch-booking follows
from the existing Dutch-book arguments for conditioning. This is because,
roughly speaking, it is more challenging to Dutch-book an agent using only
bets on $\mathcal{S}$ rather than $\mathcal{S\times L}$ (which would allow
to use the partitional framework). The other direction of the Dutch-book
theorem for forward-consistency, instead, requires a different argument,
because the payouts of a bet in a contingency must be balanced along
different continuation paths that are consistent with the same states. The
equivalence between forward consistency and Bayes plausibility is standard.

\begin{theorem}
\label{Thm:For-Cons}Consider an agent with belief system $\mu =(\mu (\cdot
|h))_{h\in \mathcal{H}}$. The following are equivalent:

\begin{enumerate}
\item $\mu $ is Bayes-plausible;

\item $\mu $ is forward-consistent;

\item the agent cannot be deterministically Dutch-booked.
\end{enumerate}
\end{theorem}

Note that forward consistency does not depend on the models the agent uses
for updating. In this sense, misspecification about the \emph{relative
probabilities of the likelihood function} is immune to deterministic
Dutch-booking. However, that is not the case for the stronger form of
misspecification where in some contingency $h$ some objectively impossible
state $s\notin S\left( h\right) $ is deemed possible by the decision maker.
This is because, by assigning an arbitrarily high positive payout $g(s|h)>>0$
to the agent in $s$ given $h$, the bookmaker can obtain a profit under some
state $s^{\prime }\in S\left( h\right) $ that is objectively possible.

Before proving the theorem, we draw the conclusions of the running example
concerning deterministic Dutch booking.

\begin{example}
As anticipated, even if we introduced an initial belief for Larry, his
beliefs would certainly be forward-consistent/Bayes-plausible, and he could
not be deterministically Dutch-booked. So, to introduce the possibility of
deterministic Dutch-booking, assume the following version of
\textquotedblleft Murphy's law\textquotedblright : the arriving bus never
goes to the fireworks location. Then, only the three contingencies $ps,sm,mp$
are possible after $h^{0}$. Suppose that, at $h^{0}$, Larry assigns
probability $2/3$ to marina and $1/6$ to each of the other two states.
However, he has uniform beliefs over the two possible states in each of the
three following contingencies. Then, his beliefs do not satisfy
Bayes-plausibility, as he assigns to marina a larger probability at $h^{0}$
than in any other future contingency. Indeed, Larry can be deterministically
Dutch-booked.\ At $h^{0}$, the bookmaker can offer him a bet that pays $%
x+\varepsilon $ for marina and $-2x$ for park and square. Then, at each $%
h\in \left\{ sm,mp\right\} $, she can offer $-x-2\varepsilon $ for marina
and $2x$ for the other possible state. (Complete the system of bets with the
null bet at $ps$.) For sufficiently small $\varepsilon $, Larry will accept.
It is easy to check that the system of bets is a deterministic Dutch book.%
\footnote{%
Note that, if we reintroduce the three contingencies where Larry learns the
state and associate them with null bets, the system of bets would no longer
be a deterministic Dutch book (consistently with what we observed in Example %
\ref{Ex:Larry3}). The reason is that the possibility to learn state marina
introduces the learning path $(h^{0},ma)$ along which Larry makes a positive
payoff.}\hfill $\triangle $
\end{example}

\bigskip

\textbf{Proof of Theorem \ref{Thm:For-Cons}.}

$\mathbf{1\Rightarrow 2)}$ This is a known fact; however, we provide formal
proof in the Appendix.

$\mathbf{2\Rightarrow 3)}$ This can be easily derived as a consequence of
Dutch-book arguments for conditionalization, and therefore a proof is
deferred to the Appendix.

$\mathbf{3\Rightarrow 1)}$ We prove the counterpositive. Suppose that $\mu $
is not Bayes-plausible. Thus, there exists a non-terminal contingency $h$
such that $\mu (\cdot |h)$ is not a convex combination of the beliefs in the
next contingencies $h^{1},...,h^{n}$. Let $A\in 
\mathbb{R}
^{\left\vert \mathcal{S}\right\vert \times n}$ be the matrix of subjective
probabilities $\mu (s|h^{j})$, and let $b\in 
\mathbb{R}
^{\left\vert \mathcal{S}\right\vert }$ be the vector of subjective
probabilities $\mu (s|h)$. So, for every transposed vector $%
x=(x_{1},...,x_{n})\in \left[ 0,1\right] ^{n}$ such that $%
\sum_{m=1}^{n}x_{m}=1$, $Ax\not=b$. Moreover, for every $x\in \left[
0,\infty \right) ^{n}$ such that $\sum_{m=1}^{n}x_{m}\not=1$, $Ax\not=b$,
because $Ax=b$ implies%
\begin{eqnarray*}
\sum_{m=1}^{n}x_{m} &=&\sum_{j=1}^{n}\left( \sum_{s\in \mathcal{S}}\mu
(s|h^{j})\right) x_{j} \\
&=&\sum_{s\in \mathcal{S}}\sum_{j=1}^{n}x_{j}\mu (s|h^{j})=\sum_{s\in 
\mathcal{S}}\mu (s|h)=1.
\end{eqnarray*}%
Thus, there is no $x\geq 0$ such that $Ax=b$. But then, by Farkas lemma,
there exists $y\in 
\mathbb{R}
^{\left\vert \mathcal{S}\right\vert }$ such that%
\begin{equation*}
A^{T}y\geq 0\text{ and }b^{T}y<0.
\end{equation*}%
For some $\varepsilon >0$, by $-b^{T}y>0$ the gamble with vector of prizes $%
-y-2\overrightarrow{\varepsilon }$ is acceptable at $h$, and by $A^{T}y\geq
0 $ the gamble with vector of prizes $y+\overrightarrow{\varepsilon }$ is
acceptable at each $h^{j}$.\ Complete the system with gambles with zero
prizes for all states in every other contingency. We obtain a deterministic
Dutch book: if one of the paths in $L\left( h\right) $ realizes (note that $%
L\left( h\right) $ contains at least some $l\in L(s)$ for some $s\in S\left(
h\right) \neq \emptyset $) the profit of the bookmaker is $\varepsilon >0$
(satisfying (\ref{Eq: strict deterministic})), otherwise it is $0$
(satisfying (\ref{Eq: DDB})).\hfill $\blacksquare $

\section{The Dutch-book theorem for complete consistency\label{Main Section}}

The Dutch-book theorem for complete consistency generalizes the insights
from the example:\ incongruities among generalized odds ratios can be
exploited for Dutch-booking and can only be displayed by a belief system
that is not completely consistent. The theorem also shows the opposite
implications: when there is no incongruity between generalized odds ratios,
the agent's belief system is completely consistent, and he cannot be
Dutch-booked. Thus, coherence among generalized odds ratios fully
characterizes complete consistency and impossibility of Dutch-booking.

\begin{theorem}
\label{Th: complete cons}Fix a belief system $\mu $. The following are
equivalent:

\begin{enumerate}
\item $\mu $ is completely coherent;

\item $\mu $ is completely consistent;

\item the agent cannot be Dutch-booked.
\end{enumerate}
\end{theorem}

Before proving the theorem, we conclude the running example.

\begin{example}
Suppose that Larry has the belief system $\hat{\mu}$ constructed before. As
we observed, $\hat{\mu}$ is not completely coherent. Thus, by Theorem \ref%
{Th: complete cons}, $\hat{\mu}$ is not completely consistent, and Larry can
be Dutch-booked. In particular, he is willing to accept Dutch book $\hat{g}$:%
\begin{equation*}
\sum_{s\in \mathcal{S}}\hat{\mu}(s|h)\hat{g}(s|h)=\frac{3}{4}\cdot 9+\frac{1%
}{4}(-10)+0\cdot 0>0\qquad \forall h\in \left\{ sm,mp,ps\right\} .
\end{equation*}%
Thus, no matter the fireworks' location, Larry ends up with equal
probability in each of the other two locations, and hence he suffers an
expected loss from this betting behavior. The lack of complete consistency
means that Larry's beliefs cannot be derived from a (lexicographic) prior
under the true DGP. This could be because Larry does not have a prior; after
all, he does not need one, as he cannot choose which bus to take. Or, it
could be that Larry updates a prior with a misspecified DGP, perceiving a
correlation between bus arrivals and the state when, in fact, there is none.
In Section \ref{Sec:Corr-Neg}, we will instead provide opposite examples of
correlation \emph{neglect}.\hfill $\triangle $
\end{example}

\bigskip

\textbf{Proof. }The acronyms DOR\ and GOR will stand for
discounted/generalized odds ratio.

\bigskip

$\mathbf{1\Rightarrow 2)}$ Note preliminarily that, by Condition 1, if a
state $s$ has a DOR\ with another state $s^{\prime }$, all the DORs between $%
s$ and $s^{\prime }$ take the same value;\footnote{%
Recall that by definition an \textquotedblleft indeterminate
DOR\textquotedblright\ between $s$ and $s^{\prime }$ is just not considered
a DOR, and likewise for GOR. We say that $s$ has a DOR\ with $s^{\prime }$
when there exists $h\in \mathcal{H}$ such that $s,s^{\prime }\in S(h)$ and $%
\mu (s|h)+\mu (s^{\prime }|h)>0$, the DOR\ between $s$ and $s^{\prime }$
being $\left( \mu (s|h)p(h|s^{\prime })\right) /\left( \mu (s^{\prime
}|h)p(h|s)\right) $. Similarly, we say that $s$ has a GOR\ with $s^{\prime }$
when it is possible to compute at least one GOR\ between $s$ and $s^{\prime
} $.} call it $o(s,s^{\prime })$.

Let $P^{1}$ be the set of all $s\in \mathcal{S}$ such that for all $%
s^{\prime }\in \mathcal{S}$, all the GORs between $s$ and $s^{\prime }$ are
different from $0$.

Let $P^{2}$ be the set of all $s\in \mathcal{S}\backslash P^{1}$ such that
for all $s^{\prime }\in \mathcal{S}\backslash P^{1}$, all the GORs between $%
s $ and $s^{\prime }$ are different from $0$.

Let $P^{3}$ be the set of all $s\in \mathcal{S}\backslash \left( P^{1}\cup
P^{2}\right) $ such that for all $s^{\prime }\in \mathcal{S}\backslash
\left( P^{1}\cup P^{2}\right) $, all the GORs between $s$ and $s^{\prime }$
are different from $0$.

And so on.

\bigskip

\begin{claim}
\label{Claim:partition}\textit{There exists }$n\geq 1$\textit{\ such that }$%
(P^{1},...,P^{n})$\textit{\ is a partition of }$\mathcal{S}$\textit{, with
non-empty elements.}
\end{claim}

\emph{Proof of the Claim.} Let $n$ be the smallest non-negative integer such
that $P^{n+1}=\emptyset $; it exists because $\mathcal{S}$ is finite and the
sets are disjoint by construction. There remains to show that $\cup
_{m=1}^{n}P^{m}=\mathcal{S}$ (which also implies that $n\geq 1$). Let $%
\widetilde{S}=\mathcal{S}\backslash \left( \cup _{m=1}^{n}P^{m}\right) $ and
suppose by contraposition that $\widetilde{S}\not=\emptyset $. The fact that 
$P^{n+1}=\emptyset $ implies that every $s\in \widetilde{S}$ has a zero GOR
with some $s^{\prime }\in \widetilde{S}$. Consider a directed graph with
nodes $\widetilde{S}$ and arrows that represent a zero GOR\ between the
start and the end node. Since every node is the start of an arrow, the graph
has a cycle. So, the product of all the (zero) GORs along the cycle is a
zero GOR\ between some $s^{1}\in \widetilde{S}$ and itself. Inverting all
the DORs in this GOR, we obtain an infinite GOR between $s^{1}$ and itself.
This violates the assumption that all the GORs\ between $s^{1}$ and itself
are identical by Condition 1.\hfill $\square $

\bigskip

\begin{claim}
\textit{There exists a LCPS }$\bar{\mu}=(\nu ^{1},...,\nu ^{n})$\textit{\
such that, for each }$m\in \left\{ 1,...,n\right\} $\textit{, (a) }$\mathrm{%
supp}\nu ^{m}=P^{m}$\textit{\ and (b), for every pair }$\left( s,s^{\prime
}\right) \in P^{m}\times P^{m}$\textit{\ which has a\ DOR }$o(s^{\prime },s)$%
\textit{, }$\nu ^{m}(s^{\prime })/\nu ^{m}(s)=o(s^{\prime },s)$.
\end{claim}

\emph{Proof of the Claim.} For each $m\in \left\{ 1,...,n\right\} $, we will
construct a probability measure $\nu ^{m}$ that satisfies (a) and (b); then, 
$(\nu ^{1},...,\nu ^{n})$ is an LCPS because, by Claim \ref{Claim:partition}%
, $(P^{1},...,P^{n})$ partitions $\mathcal{S}$.

Fix $m\in \left\{ 1,...,n\right\} $ and construct a graph with nodes $P^{m}$
and connections between nodes that have a DOR. For every connected component 
$\widetilde{P}\subseteq P^{m}$, pick a state $s$ arbitrarily, let $%
\widetilde{\nu }(s)=1$, and progressively let $\widetilde{\nu }(s^{\prime
\prime })=\widetilde{\nu }(s^{\prime })\cdot o(s^{\prime \prime },s^{\prime
})$ for every $s^{\prime \prime }\in \widetilde{P}$, where $o(s^{\prime
\prime },s^{\prime })$ is a DOR\ between $s^{\prime \prime }$ and some $%
s^{\prime }\in \widetilde{P}$ for which $\widetilde{\nu }(s^{\prime })$ was
already defined; by definition of $P^{m}$, $o(s^{\prime \prime },s^{\prime
}) $ is not zero and not infinite (otherwise $o(s^{\prime },s^{\prime \prime
})$ would be zero). So, we can obtain a probability measure $\nu ^{m}$ with
support $P^{m}$ by rescaling $\widetilde{\nu }$.

To show that $\nu ^{m}$ satisfies the desiderata, fix $s,s^{\prime }\in
P^{m} $\textit{\ }that have a\ DOR $o(s^{\prime },s)$, thus $s$ and $%
s^{\prime }$ are in the same connected component. Then, the graph has a path 
$(\bar{s}^{1},...,\bar{s}^{T})$ with $\bar{s}^{1}=s$ and $\bar{s}%
^{T}=s^{\prime }$ such that, for each $t\in \left\{ 2,...,T\right\} $, $%
\widetilde{\nu }(\bar{s}^{t})=\widetilde{\nu }(\bar{s}^{t-1})\cdot o(\bar{s}%
^{t},\bar{s}^{t-1})$, where $o(\bar{s}^{t},\bar{s}^{t-1})\in 
\mathbb{R}
^{++}$ is a DOR between $\bar{s}^{t}$ and $\bar{s}^{t-1}$. That is, there is
a path in the graph such that for every two consecutive states, the value of 
$\tilde{\nu}$ for one state has been defined by the other. (Note that, if $%
\widetilde{\nu }(\bar{s}^{t-1})$ was defined as $\widetilde{\nu }(\bar{s}%
^{t})\cdot o(\bar{s}^{t-1},\bar{s}^{t})$ and not vice versa, there is also a
DOR\ $o(\bar{s}^{t},\bar{s}^{t-1})=1/o(\bar{s}^{t-1},\bar{s}^{t})$ between $%
\bar{s}^{t}$ and $\bar{s}^{t-1}$.) Therefore%
\begin{equation*}
\frac{\nu ^{m}(s^{\prime })}{\nu ^{m}(s)}=\frac{\widetilde{\nu }(s^{\prime })%
}{\widetilde{\nu }(s)}\equiv \frac{\widetilde{\nu }(\bar{s}^{T})}{\widetilde{%
\nu }(\bar{s}^{1})}=o(\bar{s}^{T},\bar{s}^{T-1})\cdot \frac{\widetilde{\nu }(%
\bar{s}^{T-1})}{\widetilde{\nu }(\bar{s}^{1})}=...=\prod\limits_{t=2}^{T}o(%
\bar{s}^{t},\bar{s}^{t-1}).
\end{equation*}%
This is a GOR\ between $s^{\prime }$ and $s$, and it is equal to $%
o(s^{\prime },s)$ by Condition 1.\hfill $\square $

\bigskip

Now fix $h\in \mathcal{H}$. Let $m$ be the smallest $k$ such that $\nu
^{k}(S(h))>0$. We show that $\mu (\cdot |h)$ can be derived from $\nu ^{m}$
with Bayes rule. Let $\hat{\nu}^{m}$ be the probability measure derived from 
$\nu ^{m}$ with Bayes rule given $h$ (and $\left( \eta ^{s}\right) _{s\in 
\mathcal{S}}$). By property (a) of $\bar{\mu}$, $\mathrm{supp}\nu ^{m}=P^{m}$%
, thus $\mathrm{supp}\hat{\nu}^{m}=P^{m}\cap S(h)=:\hat{S}$. Moreover, for
all $s,s^{\prime }\in \hat{S}$,%
\begin{equation*}
\frac{\hat{\nu}^{m}(s)}{\hat{\nu}^{m}(s^{\prime })}=\frac{p(h|s)\nu ^{m}(s)}{%
p(h|s^{\prime })\nu ^{m}(s^{\prime })}.
\end{equation*}%
To show that $\mu (\cdot |h)$ and $\hat{\nu}^{m}$ coincide, it is enough to
show that $\mathrm{supp}\mu (\cdot |h)=\hat{S}$ and $\mu (s|h)/\mu
(s^{\prime }|h)=\hat{\nu}^{m}(s)/\hat{\nu}^{m}(s^{\prime })$ for all $%
s,s^{\prime }\in \hat{S}$.

First, we show that $\mathrm{supp}\mu (\cdot |h)=P^{m}\cap S(h)$.

Fix $s\in \mathrm{supp}\mu (\cdot |h)$. Thus, $s\in S(h)$. Hence, for every $%
k<m$, $s\not\in P^{k}$, as $\mathrm{supp}\nu ^{k}=P^{k}$ but $\nu
^{k}(S(h))=0$. Now we show that, for every $k>m$, $s\not\in P^{k}$ either,
so that $s\in P^{m}$. Suppose by contradiction that $s\in P^{k}$. As $%
\mathrm{supp}\nu ^{m}=P^{m}$ and $\nu ^{m}(S(h))>0$, there exists $\bar{s}%
\in S(h)\cap P^{m}$ such that $\nu ^{m}(\bar{s})>0$. Since $s\in \mathrm{supp%
}\mu (\cdot |h)$, $s$ has a non-zero DOR (which can be finite or infinite)
with every $s^{\prime }\in S(h)$, so in particular with $\bar{s}$. Since $%
s\not\in \cup _{i=1}^{m}P^{i}$, $s$ has a zero GOR with some $s^{\prime }\in
\cup _{i=m}^{n}P^{i}$. The two things combined imply that $\bar{s}$ has a
zero GOR with $s^{\prime }$ as well. But this contradicts $\bar{s}\in P^{m}$.

Now fix $s\not\in \mathrm{supp}\mu (\cdot |h)$. If $s\not\in S(h)$, we are
done. If $s\in S(h)$, $s\not\in P^{m}$, because $s$ has a zero DOR with
every $s^{\prime }\in \mathrm{supp}\mu (\cdot |h)$, and we have just shown
that $\mathrm{supp}\mu (\cdot |h)\subseteq P^{m}$.

To conclude the proof, fix a pair $s,s^{\prime }\in \mathrm{supp}\mu (\cdot
|h)$ and write%
\begin{equation*}
\frac{\mu (s|h)}{\mu (s^{\prime }|h)}=\frac{p(h|s)}{p(h|s^{\prime })}\frac{%
\mu (s|h)}{p(h|s)}\frac{p(h|s^{\prime })}{\mu (s^{\prime }|h)}=\frac{p(h|s)}{%
p(h|s^{\prime })}o(s,s^{\prime })=\frac{p(h|s)\nu ^{m}(s)}{p(h|s^{\prime
})\nu ^{m}(s^{\prime })},
\end{equation*}%
where $o(s,s^{\prime })$ is the DOR\ between $s$ and $s^{\prime }$ at $h$,
and the last equality follows from property (b) of $\nu ^{m}$.

\bigskip

$\mathbf{2\Rightarrow 3)}$ Let $\bar{\mu}=(\mu ^{1},...,\mu ^{n})$ be an
LCPS from which the belief system is derived. For each $m\in \left\{
1,...,n\right\} $, let $S^{m}$ denote the support of $\mu ^{m}$, and let $%
H^{m}\ $denote the set of all $h\in \mathcal{H}$ such that $\mu (\cdot |h)$
is derived from $\mu ^{m}$.

Fix a system of gambles $g=(g(\cdot |h))_{h\in \mathcal{H}}$. Suppose that $%
i $ is willing to accept $g$. We show that $g$ is not a Dutch book. Suppose
that $g\ $satisfies condition (\ref{Eq: DB1}) of a Dutch book, so that,%
\begin{equation}
\forall s\in \mathcal{S},\text{ \ \ }\sum_{h\in \mathcal{H}}p(h|s)g(s|h)\leq
0.  \label{Eq: otherwise}
\end{equation}%
We are going to show that $g(s|h)=0$ for all $s\in \mathcal{S}$ and $h\in 
\mathcal{H}$, so that $g$ does not satisfy condition (\ref{Eq: DB2}) of a
Dutch book.

The proof is by induction on $m$. Fix $m\in \left\{ 1,...,n\right\} $ and
suppose by induction that for every $k<m$, $g(s|h)=0$ for all $s\in \mathcal{%
S}$ and $h\in H^{k}$. (This is vacuously true for $m=1$.) Fix $s\in S^{m}$.
For every $k>m$ and $h\in H^{k}$, $p(h|s)=0$, otherwise $\mu (\cdot |h)$
could be derived from $\mu ^{m}$. By this and the induction hypothesis,
inequality (\ref{Eq: otherwise}) yields%
\begin{equation*}
\sum_{h\in H^{m}}p(h|s)g(s|h)\leq 0.
\end{equation*}%
So, considering that $\mu ^{m}(s^{\prime })=0$ for every $s^{\prime }\not\in
S^{m}$, we can write%
\begin{equation}
\sum_{s\in S}\mu ^{m}(s)\sum_{h\in H^{m}}p(h|s)g(s|h)\leq 0\text{.}
\label{Eq: otherwise aggr}
\end{equation}%
Since $\sum_{s^{\prime }\in \mathcal{S}}p(h|s^{\prime })\mu ^{m}(s^{\prime
})>0$ for every $h\in H^{m}$, we can rewrite (\ref{Eq: otherwise aggr})\ as%
\begin{equation}
\sum_{h\in H^{m}}\sum_{s\in \mathcal{S}}\frac{p(h|s)\mu ^{m}(s)}{%
\sum_{s^{\prime }\in \mathcal{S}}p(h|s^{\prime })\mu ^{m}(s^{\prime })}%
g(s|h)\leq 0.  \label{Eq: otherwise aggr 2}
\end{equation}%
For each $h\in H^{m}$, since the agent is willing to accept $g(\cdot |h)$,
we have%
\begin{equation}
\sum_{s\in \mathcal{S}}\mu (s|h)g(s|h)\geq 0.  \label{Eq: accept}
\end{equation}%
Since $\mu (\cdot |h)$ can be derived from $\mu ^{m}$ by Bayes rule, we can
rewrite (\ref{Eq: accept}) as%
\begin{equation}
\forall h\in H^{m},\text{ \ \ }\sum_{s\in \mathcal{S}}\frac{p(h|s)\mu ^{m}(s)%
}{\sum_{s^{\prime }\in \mathcal{S}}p(h|s^{\prime })\mu ^{m}(s^{\prime })}%
g(s|h)\geq 0.  \label{Eq: accept 2}
\end{equation}%
Equations (\ref{Eq: accept 2}) and (\ref{Eq: otherwise aggr 2}) imply%
\begin{equation*}
\forall h\in H^{m},\text{ \ \ }\sum_{s\in \mathcal{S}}\frac{p(h|s)\mu ^{m}(s)%
}{\sum_{s^{\prime }\in \mathcal{S}}p(h|s^{\prime })\mu ^{m}(s^{\prime })}%
g(s|h)=0,
\end{equation*}%
that is,%
\begin{equation*}
\sum_{s\in \mathcal{S}}\mu (s|h)g(s|h)=0.
\end{equation*}%
This means that the agent has no strict incentive to accept $g(\cdot |h)$,
therefore acceptance implies $g(s|h)=0$ for all $s\in \mathcal{S}$.

\bigskip

$\mathbf{3\Rightarrow 1)}$ It will follow as an implication of the more
general Proposition \ref{Pro:Robustness}.\hfill $\blacksquare $

\subsection{Robustness to the Bookmaker Misspecification}

One natural concern about the relevance of the previous result is that a
misspecified agent has been shown to be exploitable by a bookmaker who
perfectly knows the true data-generating process. In principle, if the
bookmaker model is also incorrect, then the proposed bet may induce some
losses for the bookmaker in some $s\in S$. We next show that this is not the
case as long as the misspecification is sufficiently small, and we quantify
what small means. To do so, we start by observing that beliefs that are not
completely coherent admit a special form of inconsistency in the generalized
discounted odd ratios.

\begin{lemma}
\label{lem:Right-Gor}If $\mu $ is not completely coherent, then there exists
a state $s$ and a GOR between $s$ and itself with value $r<1$ and where $%
s^{0},...,s^{n-1}$ are all distinct for each $k\in \left\{ i,...,n-1\right\} 
$, $h^{k}\not=h^{k+1}$.
\end{lemma}

The lemma shows that without complete coherence, the agent's belief can be
used to build a GOR that starts and ends at the same state and differs from $%
1$, the only value it could instead assume under complete coherence. This
observation motivates the following definition.

\begin{definition}
Let $r<1$ and $n\geq 2$. We say that $\mu $ is $(r,n)$\textbf{-not-coherent}
if there exists a state $s$ and a GOR between $s$ and itself with value $r<1$
and where $s^{0},...,s^{n-1}$ are all distinct for each $k\in \left\{
i,...,n-1\right\} $, $h^{k}\not=h^{k+1}$.
\end{definition}

This definition quantifies the degree of incoherence between the agent's
beliefs $\mu $ and the data generating process $p$, with smaller values of $%
r $ and $n$ indicating a more significant incoherence. Indeed, a smaller $r$
indicates that a larger departure from $1$ of the GOR between $s$ and
itself, and a smaller $n$ indicates that such a departure can be obtained by
just using a smaller number of comparisons between state-contingency pairs.
Next, we define what it means for a Dutch book to be robust to
misspecification on the bookmaker side.

\begin{definition}
Let $\beta \in \mathbb{R}_{+}$. A Dutch book $g=(g(\cdot |h))_{h\in \mathcal{%
H}}$ is a $\beta $\textbf{-robust} \textbf{Dutch book} if for every $\hat{p}$
such that 
\begin{equation*}
\frac{\hat{p}(h|s)}{\hat{p}(h^{\prime }|s)}\frac{p(h^{\prime }|s)}{p\left(
h|s\right) }<\beta \qquad \forall h,h^{\prime }\in \mathcal{H},s\in S
\end{equation*}%
we have that for all $s\in \mathcal{S}$,%
\begin{equation*}
\sum_{h\in \mathcal{H}}\hat{p}(h|s)g(s|h)\leq 0,
\end{equation*}%
and for some $s\in \mathcal{S}$,%
\begin{equation*}
\sum_{h\in \mathcal{H}}\hat{p}(h|s)g(s|h)<0.
\end{equation*}
\end{definition}

This refinement of the Dutch book concept should be interpreted as follows.
As before, the agent is endowed with beliefs $\mu $, and the bookmaker
considers the data generating process $p$. However, she no longer completely
trusts it and envisions the possibility that their understanding of how a
state translates into different consequences (i.e., $\frac{p(h^{\prime }|s)}{%
p\left( h|s\right) }$) may be incorrect to some extent. Therefore, they
require the Dutch book to lead to a payoff that is nonnegative under every
state and positive under some state, even if the true data generating
process is some different $\hat{p}$ that is sufficiently close to $p$.

The next result shows that such robustness can be achieved, and it links it
to the quantitative notion of incoherence defined above.

\begin{proposition}
\label{Pro:Robustness}Suppose that $\mu $ is $(r,n)$-not-coherent. Then for
every $\delta >0$ there exists a $\left( \sqrt[n]{\frac{1}{r}}-\delta
\right) $-robust Dutch book.
\end{proposition}

Importantly, the result shows that to achieve a Dutch book, what is key is
the comparison between the disagreement between the agent and the bookmaker,
and the misspecification of the bookmaker. If the disagreement between the
two, captured by the measure $\left( r,n\right) $ of how far are beliefs $%
\mu $ from being generated by the bookmaker model $p$ and some prior, is
sufficiently high compared to the extent of misspecification of $p$ with
respect to $\hat{p}$.

\section{Applications to Correlation Neglect\label{Sec:Corr-Neg}}

We next apply our main theorem to one of the most pervasive and studied
forms of misspecification, correlation neglect. We start by studying how the
use of different financial instruments in venture capital can be viewed as a
Dutch book due to neglect of correlation on the startup side. We then move
to see how some offers in a variant of the classical Monty Hall problem can
also be interpreted in the same way.

\paragraph{Financial Instruments}

Consider a binary-state environment, where states $\mathcal{S}=\left\{
G,B\right\} $ indicate whether a startup is good or bad. Evidence about the
startup's quality accumulates over time through a sequence of binary signals
indicating whether key milestones are achieved. That is, the set of
contingencies is $\mathcal{H=}\left\{ \left\{ \gamma ,\beta \right\}
^{n}\right\} _{n\in \left\{ 1,...,N\right\} }$, $N\geq 4$, where $\gamma $
was a good performance and $\beta $ was a bad performance in a key
milestone. The objective data generating process over contingencies is
Markov, with two key properties. First, good firms are more likely to
achieve the milestones, and second, there is persistence in the milestone
process. Regardless of whether the startup is good or bad, the probability
of achieving the second milestone is higher if the first one has been
achieved. For concreteness, the objective DGP is recursively given by%
\begin{eqnarray*}
p\left( \gamma |G\right) &=&\frac{2}{3}=p\left( \beta |B\right) \\
p\left( \left( h,\gamma ,\gamma \right) |\left( h,\gamma \right) ,G\right)
&=&\frac{3}{4}=p\left( \left( h,\beta ,\beta \right) |\left( h,\beta \right)
,B\right) \quad \forall h\in \left\{ \gamma ,\beta \right\} ^{n},n\leq N-2 \\
p\left( \left( h,\beta ,\gamma \right) |\left( h,\beta \right) ,G\right) &=&%
\frac{1}{2}=p\left( \left( h,\gamma ,\beta \right) |\left( h,\gamma \right)
,B\right) \quad \forall h\in \left\{ \gamma ,\beta \right\} ^{n},n\leq N-2%
\text{.}
\end{eqnarray*}

The startup updates its beliefs using Bayes' rule and understands that
milestones are more likely to be reached when the state is good than when it
is bad. However, the startup suffers from correlation neglect and treats
milestone achievements as i.i.d. conditional on the state. Concretely, the
belief system of the startup is given by%
\begin{equation*}
\mu \left( G|\varnothing \right) =\frac{1}{2}\text{,}
\end{equation*}%
and for all $h\in \left\{ \gamma ,\beta \right\} ^{n},n\leq N$ 
\begin{equation}
\mu \left( G|h\right) =\frac{\left( \frac{2}{3}\right) ^{|\left\{ i\in
\left\{ 1,...,n\right\} :h_{i}=\gamma \right\} |}\left( \frac{1}{3}\right)
^{|\left\{ i\in \left\{ 1,...,n\right\} :h_{i}=\beta \right\} |}}{\left( 
\frac{2}{3}\right) ^{|\left\{ i\in \left\{ 1,...,n\right\} :h_{i}=\gamma
\right\} |}\left( \frac{1}{3}\right) ^{|\left\{ i\in \left\{ 1,...,n\right\}
:h_{i}=\beta \right\} |}+\left( \frac{2}{3}\right) ^{|\left\{ i\in \left\{
1,...,n\right\} :h_{i}=\beta \right\} |}\left( \frac{1}{3}\right) ^{|\left\{
i\in \left\{ 1,...,n\right\} :h_{i}=\lambda \right\} |}}\text{.}
\label{eq:incorrect-belief}
\end{equation}%
By Theorem \ref{Thm:For-Cons}, the startup cannot be deterministically Dutch
booked, as the beliefs are derived using Bayesian updating from the
conditional probabilities 
\begin{eqnarray*}
\widetilde{p}\left( \left( h,\gamma \right) |h,G\right) &=&\frac{2}{3}\qquad
\forall h\in \left\{ \gamma ,\beta \right\} ^{n},n\leq N-1 \\
\widetilde{p}\left( \left( h,\gamma \right) |h,B\right) &=&\frac{1}{3}\qquad
\forall h\in \left\{ \gamma ,\beta \right\} ^{n},n\leq N-1\text{.}
\end{eqnarray*}

However, a venture capital offering a combination of \textquotedblleft
purchase the firm\textquotedblright\ and \textquotedblleft convertible
debt\textquotedblright\ can Dutch-book the startup by exploiting their
correlation neglect. In particular, and without loss of generality, given
the risk neutrality of the agent, suppose that ownership of a good firm is
worth $\frac{2}{3}$ in the good state and $\frac{1}{3}$ in the bad state
(that is, each state is worth its limit distribution over positive
performances). The startup initially holds ownership. The venture capitalist
offers two different contracts in contingencies of length $4$. After the
sequence $\left( \gamma ,\beta ,\beta ,\gamma \right) $, the venture
capitalist offer to buy the startup at a price of $0.51$. This corresponds
to the bet 
\begin{equation*}
g\left( G|\left( \gamma ,\beta ,\beta ,\gamma \right) \right) =0.51-\frac{2}{%
3}\text{ and }g\left( B|\left( \gamma ,\beta ,\beta ,\gamma \right) \right)
=0.51-\frac{1}{3}\text{.}
\end{equation*}%
By equation (\ref{eq:incorrect-belief}), $\mu \left( G|\left( \gamma ,\beta
,\beta ,\gamma \right) \right) =\frac{1}{2}$, and thus the startup accepts
and sells ownership, overweighting the informativeness of the two
consecutive failures.

Instead, after sequence $\left( \gamma ,\gamma ,\gamma ,\gamma \right) $,
convertible debt at a rate below the market rate is offered to the
up-and-coming startup. This below-market-rate debt amounts to a
state-independent financing advantage of $\frac{1}{100}$ to the startup, but
in the unlikely case the startup turns out to be of the bad type, the
venture capital gains control of half of the ownership of the firm. This
corresponds to the bet\footnote{%
This contract can also be approximately implemented as a transfer of
property if the fraction of milestones over a long horizon is smaller than $%
1/2$, as this is an event that has probability approaching $1$ under $B$ and 
$0$ under $G$ as the horizon grows.} 
\begin{equation*}
g\left( G|\left( \gamma ,\gamma ,\gamma ,\gamma \right) \right) =\frac{1}{100%
}\text{ and }g\left( B|\left( \gamma ,\gamma ,\gamma ,\gamma \right) \right)
=\frac{1}{100}-\frac{1}{2}\frac{1}{3}\text{.}
\end{equation*}%
By equation (\ref{eq:incorrect-belief}), $\mu \left( G|\left( \gamma ,\gamma
,\gamma ,\gamma \right) \right) =\frac{16}{17}$ and thus the startup accepts
the convertible debt, overweighting the informativeness of these successes.
Easy computations verify that this is indeed a Dutch book.\footnote{%
In state $G$, contingency $\left( \gamma ,\beta ,\beta ,\gamma \right) $ is
reached with probability $0.0416$ with a startup payoff of $-0.1566$, and $%
\left( \gamma ,\gamma ,\gamma ,\gamma \right) $ is reached with probability $%
0.28125$ and a startup payoff of $0.01$. In state $B$, contingency $\left(
\gamma ,\beta ,\beta ,\gamma \right) $ is reached with probability $0.03125$
with a startup payoff of $0.1766$, and $\left( \gamma ,\gamma ,\gamma
,\gamma \right) $ is reached with probability $0.0416$ and a startup payoff
of $0.01-\frac{1}{6}$, resulting in strictly negative expected losses in
both states.} No matter the baseline frequency of successful startups, the
venture capitalist can, on average, extract positive returns from both
types, exploiting the acquisition of good startups after mixed performance
and convertible debt issued to bad startups after an initial sequence of
successes.

We also observe that, consistently with Theorem \ref{Th: complete cons}, the
generalized odds ratios of the startup do not agree 
\begin{eqnarray*}
o\left( G,B|\left( \gamma ,\gamma ,\gamma ,\gamma \right) \right) &=&\frac{%
\mu \left( g|\left( \gamma ,\gamma ,\gamma ,\gamma \right) \right) }{p\left(
\left( \gamma ,\gamma ,\gamma ,\gamma \right) |G\right) }\frac{p\left(
\left( \gamma ,\gamma ,\gamma ,\gamma \right) |B\right) }{\mu \left(
b|\left( \gamma ,\gamma ,\gamma ,\gamma \right) \right) }=\frac{16/17}{%
0.28125}\frac{0.0416}{1/17}=2.366 \\
&>&\frac{1/2}{0.0416}\frac{0.03125}{1/2}=\frac{\mu \left( G|\left( \gamma
,\beta ,\beta ,\gamma \right) \right) }{p\left( \left( \gamma ,\beta ,\beta
,\gamma \right) |G\right) }\frac{p\left( \left( \gamma ,\beta ,\beta ,\gamma
\right) |B\right) }{\mu \left( B|\left( \gamma ,\beta ,\beta ,\gamma \right)
\right) }=o\left( G,B|\left( \gamma ,\beta ,\beta ,\gamma \right) \right) 
\text{.}
\end{eqnarray*}%
In words, the agent interprets the good type sequence $\left( \gamma ,\gamma
,\gamma ,\gamma \right) $ too positively, as it fails to account for the
persistence of performance.

This application, among other things, illustrates the difference between our
Dutch book argument and preexisting arguments against misspecification.
Typically, it is argued that some misspecifications cannot persist because,
in the long run, key statistics of the data force the agent to either reject
or incompletely trust their model. However, the form of misspecification
above would be immune to these arguments. Indeed, as the horizon grows,
i.e., $N\rightarrow \infty $, the startup beliefs concentrate on the actual
state, i.e., $\mu \left( G|h\right) $ converges to $1$ almost surely
conditional on being in state $G$, and the same holds for $\mu \left(
B|h\right) $ under $B$. Moreover, the startup will observe a fraction of
success equal to the one predicted by their incorrect model (i.e., the
stationary distribution of $\gamma $ under $g$ is $2/3$, which is the
probability of $\gamma $ under their i.i.d. model). Under essentially all
the work that studied the persistence of misspecification, these two
conditions are sufficient to guarantee that their misspecified model will
not be relaxed or changed (Gagnon-Bartsch, Rabin, and Schwartzstein 2023,
Fudenberg and Lanzani 2023, Ba 2025, He and Libgober 2025, Lanzani 2025).
Our argument against misspecification, instead, does not work at the level
of beliefs and how they evolve over time, but at the level of material
losses deriving from the cross-sectional exploitation of a population of
misspecified agents.

\paragraph{Monty Hall offers}

This is a variation of the famous Monty Hall problem, in which the initially
selected door is determined at random and not revealed to the decision
maker. There are three doors. A car is placed behind one of the doors, and a
goat is placed behind each of the other two doors. The state represents the
door with the car: $\mathcal{S}=\left\{ A,B,C\right\} $. Then, one door,
call it $d$, is drawn from a uniform distribution (under the Monty Hall
interpretation, this is the door initially assigned to the agent). Finally,
another door behind which there is a goat, call it $h\not=d$, is opened; it
is the first non-car door preceding $d$ in an $a$-$b$-$c$ cycle. So, for
instance, conditional on the state being $B$, door $a$ is opened with
probability $2/3$ (the probability that $d\in \left\{ b,c\right\} $), and
door $c$ is opened with probability $1/3$ (the probability that $d=a$). The
agent only observes which door is opened, so the possible contingencies are: 
$\mathcal{H}=\left\{ a,b,c\right\} $. Overall, conditional on each $s\in 
\mathcal{S}$, the probabilities of each $h\in \mathcal{H}$ are:%
\begin{equation*}
\begin{tabular}{|c|c|c|c|}
\hline
$p(h|s)$ & $h=a$ & $h=b$ & $h=c$ \\ \hline
$s=A$ & $0$ & $\frac{1}{3}$ & $\frac{2}{3}$ \\ \hline
$s=B$ & $\frac{2}{3}$ & $0$ & $\frac{1}{3}$ \\ \hline
$s=C$ & $\frac{1}{3}$ & $\frac{2}{3}$ & $0$ \\ \hline
\end{tabular}%
\end{equation*}%
We next consider two possible belief systems for the agent.

First, suppose that, at every $h\in \mathcal{H}$, the agent assigns equal
probability to each of the two states in $S(h)$. That is, they understand
that the car cannot be behind the open door, but they neglect the
correlation between the open door and the one containing the car, induced by
the open-in-cycle mechanism. Then, the agent can be Dutch-booked (in
objective expected terms). To see this, compute the three discounted odds
ratios:%
\begin{eqnarray*}
o(B,C|a) &=&\frac{p(a|C)}{p(a|B)}=\frac{1/3}{2/3}=\frac{1}{2}; \\
o(C,A|b) &=&\frac{p(b|A)}{p(b|C)}=\frac{1/3}{2/3}=\frac{1}{2}; \\
o(B,A|c) &=&\frac{p(c|A)}{p(c|B)}=\frac{2/3}{1/3}=2.
\end{eqnarray*}%
Since%
\begin{equation*}
o(B,C|a)\cdot o(C,A|b)\not=o(B,A|c),
\end{equation*}%
we have two different generalized odds ratios between $B$ and $A$, and thus
by Theorem \ref{Th: complete cons} the agent can be Dutch-booked. For
example, suppose that, at each $h\in \mathcal{H}$, a bookmaker offers to the
agent the following bets:%
\begin{eqnarray*}
g(C|a) &=&3,\text{ \ }g(B|a)=-2; \\
g(A|b) &=&3,\text{ \ }g(C|b)=-2; \\
g(B|c) &=&3,\text{ \ }g(A|c)=-2.
\end{eqnarray*}%
Clearly, the agent accepts all bets, and the objective expected payoff of
the bookmaker conditional on each $s\in \mathcal{S}$ is:%
\begin{eqnarray*}
-g(A|b)p(b|A)-g(A|c)p(c|A)
&=&-g(B|c)p(c|B)-g(B|a)p(a|B)=-g(C|a)p(a|C)-g(C|b)p(b|C) \\
&=&-3\frac{1}{3}+2\frac{2}{3}>0\text{.}
\end{eqnarray*}%
So, the agent is harmed by the \textquotedblleft correlation
neglect\textquotedblright\ he experiences.

Second, suppose that the agent derives his beliefs as follows. He has a
uniform prior, and, thinking of the correct updating in the original Monty
Hall problem, believes that the car is more likely to be behind the unopened
door that has not been assigned to him (i.e., not door $d$). For instance,
suppose that we are in contingency $a$ and the agent correctly computes%
\begin{eqnarray*}
\Pr \left( d=b|h=a\right) &=&\frac{\Pr (h=a|d=b)\Pr (d=b)}{\Pr (h=a|d=b)\Pr
(d=b)+\Pr (h=a|d=c)\Pr (d=c)} \\
&=&\frac{\Pr (h=a|\left( d,s\right) =\left( b,B\right) )+\Pr (h=a|\left(
d,s\right) =\left( b,C\right) )}{\left( 
\begin{array}{c}
\Pr (h=a|\left( d,s\right) =\left( b,B\right) )+\Pr (h=a|\left( d,s\right)
=\left( b,C\right) ) \\ 
+\Pr (h=a|\left( d,s\right) =\left( c,B\right) )+\Pr (h=a|\left( d,s\right)
=\left( c,C\right) )%
\end{array}%
\right) }=\frac{2}{2+1}=\frac{2}{3}.
\end{eqnarray*}%
So, he has $\mu (s=B|h=a)<1/2$, as prescribed by correct reasoning in the
(original) Monty Hall problem: the car is less likely to be behind the door
that is more likely to have been assigned to him. However, given\
contingency $a$, if the distribution over states is uniform, the car is more
likely to be behind door $b$:%
\begin{eqnarray*}
\Pr \left( s=B|h=a\right) &=&\frac{p\left( a|B\right) \Pr (s=B)}{p\left(
a|B\right) \Pr (s=B)+p\left( a|C\right) \Pr (s=C)} \\
&=&\frac{\frac{2}{3}\Pr (s=B)}{\frac{2}{3}\Pr (s=B)+\Pr (d=b)\Pr (s=C)}=%
\frac{\frac{2}{3}}{\frac{2}{3}+\frac{1}{3}}=\frac{2}{3}.
\end{eqnarray*}%
In other words, the agent perceives the opposite correlation between the
fundamental state and the observed contingency. This increases the
discrepancy between generalized odds ratios, making it easier for the
bookmaker to Dutch-book the agent.\footnote{%
To be precise, without limited liability, any discrepancy allows the
bookmaker to extract an infinite amount of money (in expectation) from the
agent. However, under limited liability, the higher discrepancy allows for a
strict increase in the bookmaker's expected profit.}

\section{Conclusion}

Dutch-book arguments have been used to motivate probabilistic sophistication
and conditioning. We introduce novel Dutch-book arguments that motivate two
forms of belief consistency in a learning framework a la Anscombe-Aumann,
with separation between subjective uncertainty (the state) and objective
uncertainty (lotteries over learning paths conditional on the state).
Bayes-plausibility is necessary and sufficient to avoid being Dutch-booked
in a deterministic sense, i.e., under each realized state and learning path.
A\ stronger form of coherence, guaranteed by deriving all beliefs from a
subjective lexicographic prior with the correct models, is necessary and
sufficient to avoid being Dutch-booked \textquotedblleft in objective
expected terms\textquotedblright , i.e., under each realized state, given
the corresponding lottery on the learning paths. The latter form of
Dutch-booking entails sure aggregate losses across a population of agents,
thereby showing the perils of heterogeneous priors across individuals and of
model misspecification even in the absence of an objective prior.

\appendix

\section{Appendix}

\subsection{Sequential games: connection with Battigalli et al. (2023)\ and
Siniscalchi (2020, 2022)}

In the context of a sequential game with complete information and perfect
recall, Battigalli et al. (2023) introduce a classification of the belief
systems of a player over the opponents' strategies. We study their
classification of belief systems as a special case of the analysis of the
main body of the paper. In a game, the moves of a player are independent of
the moves of the co-players that are yet to be observed, and the two jointly
determine the path of the game, thus the learning path. Therefore, the
assumption that every contingency is reached with the same probability under
every consistent state is satisfied in that context.\footnote{%
For concreteness, we stick to the sequential games application, but the
analysis extends seamlessly to situations where this assumption is
satisfied, for instance in situations where the an agent does not choose
what to observe but the factors that determine the observations are
independent of the state itself, as in the running example of the paper.}

Let $S_{-i}$ be the set of strategy profiles of the opponents of player $i$.
Let $H_{i}$ be the arborescence of information sets of player $i$, endowed
with the precedence relation $\prec $. Each information set $h_{i}$ is
associated with the event $S_{-i}(h_{i})\subseteq S_{-i}$ of the opponents
strategies that reach $h_{i}$, under some strategy of player $i$. A belief
system is an array $\mu _{i}=(\mu _{i}(\cdot |h_{i}))_{h_{i}\in H_{i}}$ of
probability measures over $S_{-i}$ such that $\mu
_{i}(S_{-i}(h_{i})|h_{i})=1 $ for every $h_{i}\in H_{i}$.

Taking as primitive a collection of conditioning events $\mathcal{C\subseteq 
}2^{S_{-i}}\backslash \left\{ \emptyset \right\} $, a \textbf{conditional
probability system} over $S_{-i}$ is an array of probability measures $\bar{%
\mu}_{i}=(\bar{\mu}_{i}(\cdot |C))_{C\in \mathcal{C}}$ such that $\bar{\mu}%
_{i}(C|C)=1$ for every $C\in \mathcal{C}$, and for every $D\in \mathcal{C}$
with $D\subset C$, for each $E\subseteq D$,%
\begin{equation*}
\bar{\mu}_{i}(E|C)=\bar{\mu}_{i}(E|D)\bar{\mu}_{i}(D|C).
\end{equation*}%
A conditional probability system is \textbf{complete} when it is defined
with respect to the collection $\mathcal{C=}2^{S_{-i}}\backslash \left\{
\emptyset \right\} $.

\begin{definition}[Battigalli et al., 2023]
A belief system $\mu _{i}=(\mu _{i}(\cdot |h_{i}))_{h_{i}\in H_{i}}$ over
the opponents' strategies is \textbf{forward consistent }if, for all $%
h_{i},h_{i}^{\prime }\in H_{i}$, if $h_{i}\prec h_{i}^{\prime }$, then%
\begin{equation*}
\mu _{i}(E|h_{i})=\mu _{i}(S_{-i}(h_{i}^{\prime })|h_{i})\mu
_{i}(E|h_{i}^{\prime })
\end{equation*}%
for every $E\subseteq S_{-i}(h_{i}^{\prime })$.
\end{definition}

Analogously to the general notion of Definition \ref{Def: FCBS}, forward
consistency corresponds to Bayesian updating of conjectures along the path
of play (with the inconsequential difference that consistency is not only
required between a contingency and its immediate successors). However, in
this particular setting Bayesian updating boils down to the chain rule.

\begin{definition}
A belief system $\mu _{i}=(\mu _{i}(\cdot |h_{i}))_{h_{i}\in H_{i}}$ over
the opponents\ strategies is \textbf{completely consistent} if there exists
a CCPS $\bar{\mu}_{i}=(\bar{\mu}_{i}(\cdot |C))_{C\in 2^{S_{-i}}\backslash
\left\{ \emptyset \right\} }$ such that%
\begin{equation*}
\mu _{i}(\cdot |h_{i})=\bar{\mu}_{i}(\cdot |S_{-i}(h_{i}))\qquad \forall
h_{i}\in H_{i}.
\end{equation*}
\end{definition}

The definition of complete consistency of Battigalli et al. (2023) uses
CCPS's to discipline beliefs in counterfactual contingencies. An information
set $h$ does not carry any additional information about the strategies of
the opponents besides ruling out all the strategies that do not belong to $%
S_{-i}(h)$; for this reason, it is appropriate to derive each belief by mere
conditioning in place of Bayes rule, from the most general theory that
explains the contingency.

\begin{proposition}
\label{INDEPENDENCE}For every $h_{i}\in H_{i}$, for every pair $%
s_{-i},s_{-i}^{\prime }\in S_{-i}(h_{i})$, and for every probability
distribution over player $i$'s strategies, the probability of reaching the
information set is identical under $s_{-i}$ and $s_{-i}^{\prime }$: $%
p(h_{i}|s_{-i})=p(h_{i}|s_{-i}^{\prime })$.
\end{proposition}

\textbf{Proof. }Under each $s_{-i}$, the probability of reaching $h_{i}$ is
the probability of the strategies of $i$ that, with $s_{-i}$, induce a path
that goes through $h_{i}$. So all we need to show is that such set of
strategies of $i$ is the same under all $s_{-i}\in S_{-i}(h_{i})$. By
perfect recall, all these strategies must prescribe the same moves before
reaching $h_{i}$. Suppose by contradiction that there exists a strategy of $%
i $ that induces a path that goes through $h_{i}$ with some $s_{-i}\in
S_{-i}(h_{i})$ but not with some other $s_{-i}^{\prime }\in S_{-i}(h_{i})$.
But then, all other strategies of $i$ that prescribe the same moves before $%
h_{i}$, will have the same effect: $h_{i}$ is not reached under $%
s_{-i}^{\prime }$.\ This contradicts that $s_{-i}^{\prime }$ is consistent
with $h_{i}$.\hfill $\blacksquare $

Although the use of CCPSs is appropriate in this context, one might as well
model the theories from which a player derives beliefs as a LCPS, and derive
each belief by conditioning the first theory in the list that is consistent
with the contingency; the two definitions of complete consistency are
equivalent.\footnote{%
The equivalence between LCPSs and CCPSs is well-known. We formalize and
prove this result for the analysis to be self-contained.}

\begin{proposition}
\label{LCPS=CCPS}Fix a belief system $\mu _{i}=(\mu _{i}(\cdot
|h_{i}))_{h_{i}\in H_{i}}$. The following are equivalent:

\begin{enumerate}
\item there exists a CCPS $\bar{\mu}_{i}=(\bar{\mu}_{i}(\cdot |C))_{C\in
2^{S_{-i}}\backslash \left\{ \emptyset \right\} }$ such that%
\begin{equation*}
\mu _{i}(\cdot |h_{i})=\bar{\mu}_{i}(\cdot |S_{-i}(h_{i}))\qquad \forall
h_{i}\in H_{i};
\end{equation*}

\item there exists a LCPS $\hat{\mu}_{i}=(\mu _{i}^{1},...,\mu _{i}^{n})$
such that%
\begin{equation*}
\mu _{i}(s_{-i}|h_{i})=\frac{\mu _{i}^{k}(s_{-i})}{\mu
_{i}^{k}(S_{-i}(h_{i}))}\qquad \forall h_{i}\in H_{i},\forall s_{-i}\in
S_{-i}(h_{i}),
\end{equation*}%
where $k$ is the smallest $m$ such that $\mu _{i}^{k}(S_{-i}(h_{i}))>0$.
\end{enumerate}
\end{proposition}

\textbf{Proof.}

$\mathbf{2\Rightarrow 1)}$ For each $C\in 2^{S_{-i}}\backslash \left\{
\emptyset \right\} $, derive $\bar{\mu}_{i}(\cdot |C)$ by conditioning the
first measure in the LCPS\ that assigns positive probability to $C$, so that
if $C=S_{-i}(h_{i})$ for some $h_{i}\in H_{i}$, then $\bar{\mu}_{i}(\cdot
|C)=\mu _{i}(\cdot |h_{i})$. There remains to show that $\bar{\mu}_{i}=(\bar{%
\mu}_{i}(\cdot |C))_{C\in 2^{S_{-i}}\backslash \left\{ \emptyset \right\} }$
is a CCPS. Now fix $D\subset C$ such that $\bar{\mu}_{i}(D|C)>0$. Thus, if $%
\bar{\mu}_{i}(\cdot |C)$ is derived from $\mu _{i}^{k}$, $\mu _{i}^{k}(D)>0$%
, and $\mu _{i}^{j}(D)\leq \mu _{i}^{j}(C)=0$ for every $j<k$. Therefore,
also $\bar{\mu}_{i}(\cdot |D)$ is derived from $\mu _{i}^{k}$ and the chain
rule is satisfied.

1$\Rightarrow $2) Create a decreasing chain of conditioning events $%
(C^{1},...,C^{n})$ as follows. Let $C^{1}=S_{-i}$. For each $k>1$, let $%
C^{k} $ be the union of all the $s_{-i}\in S_{-i}$ such that $\bar{\mu}%
_{i}(s_{-i}|C^{j})=0$ for every $j<k$. Let $n$ be the smallest $k$ such that 
$C^{k+1}=\emptyset $. Consider the list of measures $\hat{\mu}_{i}=(\bar{\mu}%
_{i}(\cdot |C^{1}),...,\bar{\mu}_{i}(\cdot |C^{n}))$. For each $h_{i}\in
H_{i}$, let $k$ be the smallest $j\leq n$ such that $\bar{\mu}%
_{i}(S_{-i}(h_{i})|C^{j})>0$ --- it exists by $C^{n+1}=\emptyset $. By $\bar{%
\mu}_{i}(S_{-i}(h_{i})|C^{j})=0$ for every $j<k$, it follows that $%
S_{-i}(h_{i})\subseteq C^{j}$. Hence, $\mu _{i}(\cdot |h_{i})$ can be
derived from $\bar{\mu}_{i}(\cdot |C^{k})$ by updating, and $\bar{\mu}%
_{i}(\cdot |C^{k})$ is also the first measure in $\hat{\mu}_{i}$ that gives
positive probability to $S_{-i}(h_{i})$. There remains to show that $\hat{\mu%
}_{i}$ is a LCPS. For each $s_{-i}\in S_{-i}$, by $C^{n+1}=\emptyset $, $%
\bar{\mu}_{i}(s_{-i}|C^{k})>0$ for some $k\leq n$, so $\hat{\mu}_{i}$ has
full joint support. Moreover, by $s_{-i}\in C^{k}$, $\bar{\mu}%
_{i}(s_{-i}|C^{j})=0$ for every $j<k$, and by $\bar{\mu}_{i}(s_{-i}|C^{k})>0$%
, $s_{-i}\not\in C^{j}$ for every $j>k$, which implies $\bar{\mu}%
_{i}(s_{-i}|C^{j})=0$. Hence, $\hat{\mu}_{i}$ has disjoint supports and is
therefore a LCPS.\hfill $\blacksquare $

\bigskip

Proposition \ref{LCPS=CCPS} implies that the Dutch-book theorem also applies
to belief consistency as it was defined for sequential games. With this, the
characterization of complete consistency with the coherence rule among
discounted odds ratios can be rewritten using traditional odds ratios in
place of discounted odds ratios, because in this context the two coincide by
Proposition \ref{INDEPENDENCE}.

\begin{definition}
Given a belief system $(\mu _{i}(\cdot |h))_{h\in H_{i}}$, an information
set $h$, and a pair $s_{-i},s_{-i}^{\prime }\in S_{-i}(h)$, let%
\begin{equation*}
o^{\ast }(s_{-i},s_{-i}^{\prime }|h)=\frac{\mu _{i}(s_{-i}|h)}{\mu
_{i}(s_{-i}^{\prime }|h)}
\end{equation*}%
denote the odds ratio between $s_{-i}$ and $s_{-i}^{\prime }$ at $h$,
provided that it is not indeterminate. A \textbf{simple} \textbf{generalized
odds ratio} between $s_{-i}$ and $s_{-i}^{\prime }$ is the product of a
finite concatenation of odds ratios%
\begin{equation*}
o^{\ast }(s_{-i},s_{-i}^{1}|h^{1})\cdot o^{\ast
}(s_{-i}^{1},s_{-i}^{2}|h^{2})\cdot ....\cdot o^{\ast
}(s_{-i}^{n-1},s_{-i}^{\prime }|h^{n})
\end{equation*}%
that is not indeterminate.
\end{definition}

The following is a corollary of the Dutch-book theorem for complete
consistency and of the last two propositions.

\begin{corollary}
\label{Cor: complete cons}Fix a belief system $\mu _{i}=(\mu _{i}(\cdot
|h_{i}))_{h_{i}\in H_{i}}$. The following are equivalent:

\begin{enumerate}
\item the belief system is completely consistent;

\item for every $s_{-i},s_{-i}^{\prime }\in S_{-i}$, all the simple
generalized odds ratios between $s_{-i}$ and $s_{-i}^{\prime }$ are
identical.
\end{enumerate}
\end{corollary}

Siniscalchi (2020) (the working paper of Siniscalchi 2022) comes to a
similar characterization of complete consistency starting from a different
idea. Consider a belief system that is isomorphic to a conditional
probability system defined over the collection of observable events.\ Such
belief system thus satisfies the chain rule between contingencies that
correspond to nested observations, but the chain rule may have little bite
when the observable events are not nested, and the corresponding beliefs are
not disciplined by the preliminary observation of a larger event. Therefore,
Siniscalchi (2020) generalizes the chain rule to apply also between
overlapping, but non-nested observable events. We conclude this section by
showing the convergence between Siniscalchi's approach and ours.

\bigskip

Call \textquotedblleft simple generalized self-odds ratio\textquotedblright\
a simple generalized odds ratio between $s_{-i}$ and itself.

\begin{remark}
\label{Remark: cycle}All the simple generalized odds ratios of each pair are
identical if and only if all simple generalized self-odds ratios are $1$.
\end{remark}

\textbf{Proof. }

\textbf{Only if} Fix a simple generalized self-odds ratio. Break it into two
and invert one of the two halves. We have two concatenations of odds ratios
that start and end with the same element. Thus, the two corresponding simple
generalized odds ratios are identical by assumption. Therefore, the simple
generalized self-odds ratio is $1$.

\textbf{If }Fix a pair and two concatenations of odds ratios of this pair
where either no odds ratio is infinite or there is no zero. If both contain
a zero, the corresponding simple generalized odds ratios are both zero. If
they both contain an infinite, they are both infinite. If one contains an
infinite and the other contains a zero, the simple generalized self-odds
ratio obtained by inverting one of the two is zero or infinite, which
violates the assumption. Otherwise, the simple generalized self-odds ratio
(in one direction) is finite. By assumption, it is $1$. Therefore the two
simple generalized odds ratios are identical.\hfill $\blacksquare $

\begin{definition}[Siniscalchi 2020]
A belief system $(\mu ^{i}(\cdot |h_{i}))_{h_{i}\in H_{i}}$ is consistent if
for every $(h_{i}^{1},...,h_{i}^{n})\in H_{i}^{n}$ and $E\subseteq
S_{-i}(h_{i}^{1})\cap S_{-i}(h_{i}^{n})$,%
\begin{equation}
\mu ^{i}(E|h_{i}^{1})\prod_{m=1}^{n-1}\mu ^{i}(S_{-i}(h_{i}^{m})\cap
S_{-i}(h_{i}^{m+1})|h_{i}^{m+1})=\mu ^{i}(E|h_{i}^{n})\prod_{m=1}^{n-1}\mu
^{i}(S_{-i}(h_{i}^{m})\cap S_{-i}(h_{i}^{m+1})|h_{i}^{m}).
\label{Eq: Marciano}
\end{equation}
\end{definition}

\begin{theorem}
A belief system $(\mu ^{i}(\cdot |h_{i}))_{h_{i}\in H_{i}}$ is consistent if
and only if it is completely consistent.
\end{theorem}

\textbf{Proof. }We will use the characterization of complete consistency of
Corollary \ref{Cor: complete cons}.

\textbf{If} Fix $(h_{i}^{1},...,h_{i}^{n})\in H_{i}^{n}$ and $E\subseteq
S_{-i}(h_{i}^{1})\cap S_{-i}(h_{i}^{n})$. If both sides of equation (\ref%
{Eq: Marciano}) are zero, consistency holds. Otherwise, suppose without loss
of generality that the right-hand side is not zero, so that equation (\ref%
{Eq: Marciano}) can be rewritten as%
\begin{equation}
\frac{\mu ^{i}(E|h_{i}^{1})}{\mu ^{i}(S_{-i}(h_{i}^{1})\cap
S_{-i}(h_{i}^{2})|h_{i}^{1})}\cdot ....\cdot \frac{\mu
^{i}(S_{-i}(h_{i}^{n-1})\cap S_{-i}(h_{i}^{n})|h_{i}^{n})}{\mu
^{i}(E|h_{i}^{n})}=1.  \label{Eq: Emarciano}
\end{equation}%
Note that equation (\ref{Eq: Emarciano}) holds if for every $s_{-i}\in E$
such that the ratio $\mu ^{i}(s_{-i}|h_{i}^{1})/\mu ^{i}(s_{-i}|h_{i}^{n})$
is not indeterminate,%
\begin{equation}
\frac{\mu ^{i}(s_{-i}|h_{i}^{1})}{\mu ^{i}(S_{-i}(h_{i}^{1})\cap
S_{-i}(h_{i}^{2})|h_{i}^{1})}\cdot ....\cdot \frac{\mu
^{i}(S_{-i}(h_{i}^{n-1})\cap S_{-i}(h_{i}^{n})|h_{i}^{n})}{\mu
^{i}(s_{-i}|h_{i}^{n})}=1,  \label{eq: Emarliano}
\end{equation}%
because this means that all the non-indeterminate ratios $\mu
^{i}(s_{-i}|h_{i}^{1})/\mu ^{i}(s_{-i}|h_{i}^{n})$ are identical, which
implies that the ratio $\mu ^{i}(E|h_{i}^{1})/\mu ^{i}(E|h_{i}^{n})$ takes
the same value too. We are going to show that for every $m\in \left\{
2,...,n\right\} $, there is $s_{-i}^{m}\in S_{-i}(h_{i}^{m-1})\cap
S_{-i}(h_{i}^{m})$ such that%
\begin{equation*}
\frac{\mu ^{i}(S_{-i}(h_{i}^{m-1})\cap S_{-i}(h_{i}^{m})|h_{i}^{m})}{\mu
^{i}(S_{-i}(h_{i}^{m-1})\cap S_{-i}(h_{i}^{m})|h_{i}^{m-1})}=\frac{\mu
^{i}(s_{-i}^{m}|h_{i}^{m})}{\mu ^{i}(s_{-i}^{m}|h_{i}^{m-1})};
\end{equation*}%
then, we can rewrite equation (\ref{eq: Emarliano}) as%
\begin{equation}
\frac{\mu ^{i}(s_{-i}|h_{i}^{1})}{\mu ^{i}(s_{-i}^{2}|h_{i}^{1})}\cdot
....\cdot \frac{\mu ^{i}(s_{-i}^{n}|h_{i}^{n})}{\mu ^{i}(s_{-i}|h_{i}^{n})}%
=1,  \label{Eq: Emirliano}
\end{equation}%
which is true by complete consistency (cf. Remark \ref{Remark: cycle}). Let $%
s_{-i}^{m}$ be any $s_{-i}\in S_{-i}(h_{i}^{m-1})\cap S_{-i}(h_{i}^{m})$
such that $\mu ^{i}(s_{-i}^{m}|h_{i}^{m-1})>0$; one exists by the fact that
the right-hand side of (\ref{Eq: Marciano}) is not zero. By complete
consistency and Corollary \ref{Cor: complete cons}, for each $s_{-i}^{\prime
}\in S_{-i}(h_{i}^{m-1})\cap S_{-i}(h_{i}^{m})$, we have%
\begin{equation*}
\frac{\mu ^{i}(s_{-i}^{m}|h_{i}^{m})}{\mu ^{i}(s_{-i}^{\prime }|h_{i}^{m})}=%
\frac{\mu ^{i}(s_{-i}^{m}|h_{i}^{m-1})}{\mu ^{i}(s_{-i}^{\prime
}|h_{i}^{m-1})}.
\end{equation*}%
With some algebra, it is easy to show that then%
\begin{equation*}
\frac{\mu ^{i}(s_{-i}^{m}|h_{i}^{m})}{\mu ^{i}(S_{-i}(h_{i}^{m-1})\cap
S_{-i}(h_{i}^{m})|h_{i}^{m})}=\frac{\mu ^{i}(s_{-i}^{m}|h_{i}^{m-1})}{\mu
^{i}(S_{-i}(h_{i}^{m-1})\cap S_{-i}(h_{i}^{m-1})|h_{i}^{m-1})},
\end{equation*}%
as we wanted to show.

\textbf{Only if} By Remark \ref{Remark: cycle} it is enough to show that for
each $s_{-i},\in S_{-i}$, for every simple generalized self-odds ratio,
equation (\ref{Eq: Emirliano}) holds. Without loss of generality, we can
assume that the simple generalized self-odds ratio is not infinite (because
an infinite one can be inverted and becomes zero), so that no denominator in
equation (\ref{Eq: Emirliano}) is zero. For each $m\in \left\{
2,...,n\right\} $, by consistency, equation (\ref{Eq: Marciano}) for $%
E=\left\{ s_{-i}^{m}\right\} $ and $\left( h_{i}^{1},...,h_{i}^{n}\right)
=(h_{i}^{m-1},h_{i}^{m})$ yields%
\begin{equation*}
\frac{\mu ^{i}(s_{-i}^{m}|h_{i}^{m})}{\mu ^{i}(s_{-i}^{m}|h_{i}^{m-1})}=%
\frac{\mu ^{i}(S_{-i}(h_{i}^{m-1})\cap S_{-i}(h_{i}^{m})|h_{i}^{m})}{\mu
^{i}(S_{-i}(h_{i}^{m-1})\cap S_{-i}(h_{i}^{m-1})|h_{i}^{m-1})}.
\end{equation*}%
Then, substituting in (\ref{Eq: Emirliano}), we obtain equation (\ref{eq:
Emarliano}), which coincides with equation (\ref{Eq: Emarciano}) for $%
E=\left\{ s_{-i}\right\} $, and thus holds by consistency.

\bigskip

\subsection{Completion of the proofs of the theorems}

\textbf{Proof of Theorem 1:\ }

$\mathbf{1\Rightarrow 2)}$ Fix a non-terminal contingency $h$ and let $%
h^{1},...,h^{n}$ be the next contingencies. Let $\alpha _{1},...,\alpha _{n}$
denote the weights that verify the definition of Bayes-plausibility. We
construct probability distributions $(\widetilde{p}(h^{m}|h,s))_{m=1}^{n}$
of reaching each $h^{m}$ given $h$ and each $s\in S(h)$. Fix $m\in \left\{
1,...,n\right\} $. If $\alpha _{m}=0$, let $\widetilde{p}(h^{m}|h,s)=0$ for
all $s\in \mathcal{S}$. Suppose now that $\alpha _{m}>0$. For each $s\in
S(h) $, let%
\begin{equation*}
\widetilde{p}(h^{m}|h,s)=\left\{ 
\begin{tabular}{ll}
$\frac{\mu (s|h^{m})\alpha _{m}}{\mu (s|h)}$ & $\text{if }s\in \mathrm{supp}%
\mu (\cdot |h)$ \\ 
$\alpha ^{m}$ & otherwise.%
\end{tabular}%
\right.
\end{equation*}%
For each $s\not\in \mathrm{supp}\mu (\cdot |h)$, we have%
\begin{equation*}
\sum_{m=1}^{n}\widetilde{p}(h^{m}|h,s)=\sum_{m=1}^{n}\alpha _{m}=1.
\end{equation*}%
For each $s\in \mathrm{supp}\mu (\cdot |h)$, we have%
\begin{equation*}
\sum_{m=1}^{n}\widetilde{p}(h^{m}|h,s)=\frac{\sum_{m=1}^{n}\mu
(s|h^{m})\alpha _{m}}{\mu (s|h)}=1,
\end{equation*}%
where the last equality follows from Bayes-plausibility.

Now fix $m\in \left\{ 1,...,n\right\} $. If $\alpha _{m}=0$, then%
\begin{equation*}
\sum_{s\in S(h)}\widetilde{p}(h^{m}|h,s)=0.
\end{equation*}%
If $\alpha _{m}>0$, by Bayes-plausibility%
\begin{equation}
\mathrm{supp}\mu (\cdot |h^{m})\subseteq \mathrm{supp}\mu (\cdot |h).
\label{Eq: BC}
\end{equation}%
So we have%
\begin{equation*}
\frac{\widetilde{p}(h^{m}|h,s)\mu (s|h)}{\sum_{s^{\prime }\in S(h)}%
\widetilde{p}(h^{m}|h,s^{\prime })\mu (s^{\prime }|h)}=\frac{\frac{\mu
(s|h^{m})\alpha _{m}}{\mu (s|h)}\mu (s|h)}{\alpha _{m}\sum_{s^{\prime }\in 
\mathrm{supp}\mu (\cdot |h)}\mu (s^{\prime }|h^{m})}=\mu (s|h^{m}),\qquad
\forall s\in S\left( h\right)
\end{equation*}%
where the last equality follows from (\ref{Eq: BC}) and 
\begin{equation*}
\frac{\widetilde{p}(h^{m}|h,s)\mu (s|h)}{\sum_{s^{\prime }\in S(h)}%
\widetilde{p}(h^{m}|h,s^{\prime })\mu (s^{\prime }|h)}=0=\mu (s|h^{m})\qquad
\forall s\notin S\left( h\right) \text{.}
\end{equation*}%
Thus, $\mu (\cdot |h^{m})$ can be derived with Bayes rule from $\mu (\cdot
|h)$ using $(\widetilde{p}(h^{m}|h,s))_{s\in S(h)}$ whenever possible.

$\mathbf{2\Rightarrow 3)}$ Fix a system of bets that the agent accepts. We
show that it is not a deterministic Dutch book. Fix an initial contingency $%
h $. Along the paths $\bar{L}$ that start from $h$ and are consistent with
some $s\in $\textrm{supp}$\mu (\cdot |h)$\textrm{, }forward-consistency is
equivalent to updating by conditioning a joint belief over states and paths
with marginal $\mu (\cdot |h)$ over the states. Thus, the Dutch-book
argument for conditioning (Teller, 1973; Lewis, 1999) guarantees that the
bookmaker either makes zero profit along every $l\in \bar{L}$, or negative
profit along some $\bar{l}\in \bar{L}$, under some state $\bar{s}$ that is
consistent with $\bar{l}$.\footnote{%
This applies to betting on the state-path pair, and hence, as a subcase,
also to our case of betting on the state only.} In the second case, the
system is not a deterministic Dutch book, as it violates condition of
equation (\ref{Eq: DDB}) with $s=\bar{s}$ and $l=\bar{l}$. The first case is
impossible: since acceptance (given the tie-breaking rule) guarantees that
the agent expects a positive payoff under every bet along the paths in $\bar{%
L}$, one could lower the payouts to the agent, preserve acceptance, and give
positive profits to the bookmaker along all the paths, contradicting the
Dutch-book argument for conditioning.

\textbf{Proof of Lemma \ref{lem:Right-Gor}. }Suppose that there exist two
different GORs between some $s$ and $s^{\prime }$. Then, inverting all the
DORs in the second GOR and multiplying by the first GOR,\footnote{%
Since the two GORs are different, they cannot be both infinite or both zero,
therefore after inverting one of the two their product is not indeterminate.}
we obtain a new GOR between $s$ and itself, which is not $1$; write it as%
\begin{equation*}
r:=\prod_{i=1}^{n}o(s^{i-1},s^{i}|h^{i}),
\end{equation*}%
where $n\geq 2$, $s^{0}=s^{n}=s$, and for each $k\in \left\{ 1,...,n\right\} 
$,%
\begin{equation}
o(s^{k-1},s^{k}|h^{k})=\frac{p(h^{k}|s^{k})}{p(h^{k}|s^{k-1})}\frac{\mu
(s^{k-1}|h^{k})}{\mu (s^{k}|h^{k})}.  \label{Eq: chain}
\end{equation}%
Without loss of generality, we can suppose that:

\begin{description}
\item[-] $s^{0},...,s^{n-1}$ are all distinct, because if $s^{j}=s^{k}$,
either $r^{\prime }=o(s^{j},s^{j+1}|h^{j})\cdot ...\cdot
o(s^{k-1},s^{k}|h^{k})$ is a new GOR between $s^{j}$ and itself that is not $%
1$, or $r^{\prime \prime }=r/r^{\prime }$ is a new GOR between $s$ and
itself that is not $1$;

\item[-] for each $k=i,...,n-1$, $h^{k}\not=h^{k+1}$, because if $%
h^{k}=h^{k+1}=h$, we have $o(s^{k-1},s^{k}|h^{k})\cdot
o(s^{k},s^{k+1}|h^{k+1})=o(s^{k-1},s^{k+1}|h)$ and we can substitute the
left-hand side with the right-hand side in the GOR (eliminating state $s^{k}$%
) --- likewise, we can suppose $h^{n}\not=h^{1}$;

\item[-] $r<1$, because if a GOR\ is larger than $1$, inverting all the DORs
we obtain a GOR smaller than $1$ (still between $s$ and itself).\hfill $%
\blacksquare $
\end{description}

\bigskip

\textbf{Proof of Proposition \ref{Pro:Robustness}.\ }Consider the GOR given
by the definition of $(r,n)$-not-coherent. We fix a $\delta >0$ and use this
GOR to construct a $\left( \sqrt[n]{\frac{1}{r}}-\delta \right) $-robust
Dutch book. Fix $\varepsilon >0$ and let $\alpha =\sqrt[n]{\frac{1}{r}}$.
Define $g(s^{0}|h^{1})$ and $g(s^{1}|h^{1})$ as:%
\begin{eqnarray*}
g(s^{0}|h^{1}) &=&-1, \\
g(s^{1}|h^{1}) &=&-g(s^{0}|h^{1})\frac{\mu (s^{0}|h^{1})}{\mu (s^{1}|h^{1})}%
+\varepsilon =\frac{\mu (s^{0}|h^{1})}{\mu (s^{1}|h^{1})}+\varepsilon .
\end{eqnarray*}%
Recursively, for each $m\in \left\{ 2,...,n\right\} $ define $%
g(s^{m-1}|h^{m})$ and $g(s^{m}|h^{m})$ as:\footnote{%
Since $h^{m}\not=h^{m-1}$, we are not redefining $g(s^{m-1}|h^{m-1})$.}%
\begin{eqnarray}
g(s^{m-1}|h^{m}) &=&-g(s^{m-1}|h^{m-1})\frac{p(h^{m-1}|s^{m-1})}{%
p(h^{m}|s^{m-1})}\sqrt[n]{\frac{1}{r}},  \label{Eq: Dutch} \\
g(s^{m}|h^{m}) &=&-g(s^{m-1}|h^{m})\frac{\mu (s^{m-1}|h^{m})}{\mu
(s^{m}|h^{m})}+\varepsilon .  \label{Eq: accepting}
\end{eqnarray}%
Complete the system of bets with $g(\widetilde{s}|h)=0$ wherever $g(%
\widetilde{s}|h)$ is not already defined.

For each $m\in \left\{ 1,...,n\right\} $, the agent is willing to accept $%
g(\cdot |h^{m})$ because%
\begin{eqnarray*}
&&\mu (s^{m-1}|h^{m})g(s^{m-1}|h^{m})+\mu (s^{m}|h^{m})g(s^{m}|h^{m}) \\
&=&\mu (s^{m-1}|h^{m})g(s^{m-1}|h^{m})-g(s^{m-1}|h^{m})\mu
(s^{m-1}|h^{m})+\varepsilon \mu (s^{m}|h^{m})=\varepsilon \mu (s^{m}|h^{m})>0%
\text{,}
\end{eqnarray*}%
where $\mu (s^{m}|h^{m})>0$ because $r\not=\infty $.\footnote{%
If $h^{m}=h^{k}=h$ for some $k\not=m$, the same inequality holds for the
other involved states, so $g(\cdot |h)$ is accepted.} Thus, the agent is
willing to accept $g=(g(\cdot |h))_{h\in \mathcal{H}}$.

Let $\hat{p}$ be such that $\max_{h,h^{\prime }\in \mathcal{H},s\in S}\frac{%
\hat{p}(h|s)}{\hat{p}(h^{\prime }|s)}\frac{p(h^{\prime }|s)}{p\left(
h|s\right) }<\left( \sqrt[n]{\frac{1}{r}}-\delta \right) $. For each $m\in
\left\{ 1,...,n-1\right\} $ (recall that $s^{n}=s^{1}$), since $%
s^{m}\not=s^{m^{\prime }}$ for every $m^{\prime }\not=m$, we have%
\begin{eqnarray*}
\sum_{h\in \mathcal{H}}\hat{p}(h|s^{m})g(s^{m}|h) &=&\hat{p}%
(h^{m}|s^{m})g(s^{m}|h^{m})+\hat{p}(h^{m+1}|s^{m})g(s^{m}|h^{m+1}) \\
&=&\hat{p}(h^{m}|s^{m})g(s^{m}|h^{m})-\hat{p}(h^{m+1}|s^{m})g(s^{m}|h^{m})%
\frac{p(h^{m}|s^{m})}{p(h^{m+1}|s^{m})}\sqrt[n]{\frac{1}{r}}<0,
\end{eqnarray*}%
where $p(h^{m+1}|s^{m})>0$ because $s^{m}\in S(h^{m+1})$ (DORs in a
contingency are only defined between consistent states). Thus, $s^{m}$
verifies the strict inequality condition of a $\left( \sqrt[n]{\frac{1}{r}}%
-\delta \right) $-robust Dutch book. Every $\widetilde{s}\in \mathcal{S}%
\backslash \left\{ s^{1},...,s^{n-1}\right\} $ trivially satisfies the weak
inequality condition. Finally, we need to show that, for some $\varepsilon
>0 $, $s$ satisfies the weak inequality condition, so that $g$ is a Dutch
book. First observe that for every $\hat{p}$ such that $\max_{h,h^{\prime
}\in \mathcal{H},s\in S}\frac{\hat{p}(h|s)}{\hat{p}(h^{\prime }|s)}\frac{%
p(h^{\prime }|s)}{p\left( h|s\right) }<\left( \sqrt[n]{\frac{1}{r}}-\delta
\right) $, we have%
\begin{eqnarray*}
p(h^{1}|s)g(s|h^{1})+p(h^{n}|s)g(s|h^{n})\left( \sqrt[n]{\frac{1}{r}}-\delta
\right) &\leq &0\iff g(s|h^{1})+\frac{p(h^{n}|s)}{p(h^{1}|s)}%
g(s|h^{n})\left( \sqrt[n]{\frac{1}{r}}-\delta \right) \leq 0 \\
&\iff &g(s|h^{1})+\frac{p(h^{n}|s)/\hat{p}(h^{n}|s)}{p(h^{1}|s)/\hat{p}%
(h^{1}|s)}\frac{\hat{p}(h^{n}|s)}{\hat{p}(h^{1}|s)}g(s|h^{n})\left( \sqrt[n]{%
\frac{1}{r}}-\delta \right) \leq 0 \\
&\implies &g(s|h^{1})+\frac{\hat{p}(h^{n}|s)}{\hat{p}(h^{1}|s)}%
g(s|h^{n})\leq 0 \\
&\iff &\hat{p}(h^{1}|s)g(s|h^{1})+\hat{p}(h^{n}|s)g(s|h^{n})\leq 0 \\
&\iff &\sum_{h\in \mathcal{H}}\hat{p}(h|s)g(s|h)\leq 0\text{.}
\end{eqnarray*}%
Note that, for $\varepsilon =0$,%
\begin{eqnarray*}
&&p(h^{1}|s)g(s|h^{1})+p(h^{n}|s)g(s|h^{n})\left( \sqrt[n]{\frac{1}{r}}%
-\delta \right) \\
&=&-p(h^{1}|s)-p(h^{n}|s)g(s^{n-1}|h^{n})\frac{\mu (s^{n-1}|h^{n})}{\mu
(s|h^{n})}\left( \sqrt[n]{\frac{1}{r}}-\delta \right) \\
&=&-p(h^{1}|s)+p(h^{n}|s)g(s^{n-1}|h^{n-1})\frac{p(h^{n-1}|s^{n-1})}{%
p(h^{n}|s^{n-1})}\sqrt[n]{\frac{1}{r}}\frac{\mu (s^{n-1}|h^{n})}{\mu
(s|h^{n})}\left( \sqrt[n]{\frac{1}{r}}-\delta \right) \\
&=&-p(h^{1}|s)+p(h^{n-1}|s^{n-1})g(s^{n-1}|h^{n-1})\sqrt[n]{\frac{1}{r}}%
o(s^{n-1},s|h^{n})\left( \sqrt[n]{\frac{1}{r}}-\delta \right) \\
&=&-p(h^{1}|s)-\beta p(h^{n-1}|s^{n-1})g(s^{n-2}|h^{n-1})\frac{\mu
(s^{n-2}|h^{n-1})}{\mu (s^{n-1}|h^{n-1})}\sqrt[n]{\frac{1}{r}}%
o(s^{n-1},s|h^{n})\left( \sqrt[n]{\frac{1}{r}}-\delta \right) \\
&=&-p(h^{1}|s)+p(h^{n-1}|s^{n-1})g(s^{n-2}|h^{n-2})\frac{p(h^{n-2}|s^{n-2})}{%
p(h^{n-1}|s^{n-2})}\left( \sqrt[n]{\frac{1}{r}}\right) ^{2}\frac{\mu
(s^{n-2}|h^{n-1})}{\mu (s^{n-1}|h^{n-1})}o(s^{n-1},s|h^{n})\left( \sqrt[n]{%
\frac{1}{r}}-\delta \right) \\
&=&-p(h^{1}|s)+p(h^{n-2}|s^{n-2})g(s^{n-2}|h^{n-2})\left( \sqrt[n]{\frac{1}{r%
}}\right) ^{2}o(s^{n-2},s^{n-1}|h^{n-1})o(s^{n-1},s|h^{n})\left( \sqrt[n]{%
\frac{1}{r}}-\delta \right) \\
&=&... \\
&=&-p(h^{1}|s)-p(h^{1}|s^{1})g(s^{0}|h^{1})\frac{\mu (s|h^{1})}{\mu
(s^{1}|h^{1})}\left( \sqrt[n]{\frac{1}{r}}\right)
^{n-1}o(s^{1},s^{2}|h^{2})\cdot ...\cdot
o(s^{n-2},s^{n-1}|h^{n-1})o(s^{n-1},s|h)\left( \sqrt[n]{\frac{1}{r}}-\delta
\right) \\
&=&-p(h^{1}|s)+p(h^{1}|s)r\left( \sqrt[n]{\frac{1}{r}}\right) ^{n-1}\left( 
\sqrt[n]{\frac{1}{r}}-\delta \right) =-p(h^{1}|s)+p(h^{1}|s)r\left( \sqrt[n]{%
\frac{1}{r}}\right) ^{n}\frac{\left( \sqrt[n]{\frac{1}{r}}-\delta \right) }{%
\sqrt[n]{\frac{1}{r}}} \\
&\leq &-p(h^{1}|s)+p(h^{1}|s)\frac{\left( \sqrt[n]{\frac{1}{r}}-\delta
\right) }{\sqrt[n]{\frac{1}{r}}}<0\text{,}
\end{eqnarray*}%
where the second equality uses (\ref{Eq: accepting}), the third (\ref{Eq:
Dutch}), the fourth (\ref{Eq: chain}), and so on. By continuity, the strict
inequality is preserved for sufficiently small $\varepsilon >0$.\hfill $%
\blacksquare $

\end{document}